\documentclass[12pt]{article}
\pdfoutput=1
\usepackage[centertags]{amsmath}
\allowdisplaybreaks[1]
\usepackage[square, comma, sort&compress,numbers]{natbib}
\usepackage{array,multirow}
\numberwithin{equation}{section}
\usepackage{amssymb}
\usepackage{graphicx}
\usepackage{color}
\usepackage{relsize}

\usepackage{epsfig}

\def\Or[#1]{{\text{O}}\left({#1}\right)}
\def\dotl[#1,#2]{\left\langle #1, #2 \right\rangle}
\def\dotlb[#1,#2]{[ #1, #2 ]}
\def\dotp[#1,#2]{(#1) \cdot (#2)}
\def\aff[#1,#2]{\hat{#1}(#2)}
\def\n4sym{{\cal N}=4 SYM}
\def\>{\rangle}
\def\<{\langle}
\def\weight[#1,#2,#3]{\{(#1),#2,#3\}}
\def\ads[#1]{$\text{AdS}_{#1}$}

\newcommand{\ba}{\begin{eqnarray}}
\newcommand{\ea}{\end{eqnarray}}
\newcommand{\be}{\begin{eqnarray}}
\newcommand{\ee}{\end{eqnarray}}
\newcommand{\bq}{\begin{equation}}
\newcommand{\eq}{\end{equation}}
\newcommand{\benn}{\begin{equation*}}
\newcommand{\eenn}{\end{equation*}}
\newcommand{\bi}{\begin{itemize}}  
\newcommand{\ei}{\end{itemize}}

\newcommand{\CI}{{\cal I}}
\newcommand{\CL}{{\cal L}}
\newcommand{\CJ}{{\cal J}}

\newcommand{\CN}{{\cal N}}
\newcommand{\CO}{{\cal O}}
\newcommand{\CP}{{\cal P}}

\newcommand{\CT}{{\cal T}}
\newcommand{\CV}{{\cal V}}
\newcommand{\nn}{\nonumber}

\newcommand\oo\infty
\newcommand\s\sigma
\newcommand\de\delta
\newcommand\De\Delta

\newcommand\f\phi
\newcommand\g\gamma
\newcommand\x\times

\newcommand{\Ocal}{{\cal O}}

\newcommand{\Vcal}{{\cal V}}

\usepackage{jheppub}

\makeatletter
\def\@fpheader{\vspace{-.1cm}}
\makeatother

\title{Degenerate Operators and the $1/c$ Expansion: Lorentzian Resummations, High Order Computations, and Super-Virasoro Blocks}

\author[a]{Hongbin Chen,}
\author[b]{A.\ Liam Fitzpatrick,}
\author[a]{Jared Kaplan,}
\author[a]{Daliang Li,}
\author[a]{and Junpu Wang}
\affiliation[a]{Department of Physics and Astronomy,  Johns Hopkins University, \\
Charles Street, Baltimore, MD 21218, U.S.A.}
\affiliation[b]{Department of Physics, Boston University, \\
Commonwealth Avenue, Boston, MA 02215, U.S.A.}

\abstract{ 
One can obtain exact information about Virasoro conformal blocks by analytically continuing the correlators of degenerate operators.  We argued in recent work that this technique can be used to explicitly resolve information loss problems in AdS$_3$/CFT$_2$.  In this paper we use the technique to perform calculations in the small $1/c \propto G_N$ expansion:  (1) we prove the all-orders resummation of logarithmic factors $\propto \frac{1}{c} \log z$ in the Lorentzian regime, demonstrating that $1/c$ corrections directly shift Lyapunov exponents associated with chaos, as claimed in prior work, (2) we perform another all-orders resummation in the limit of large $c$ with fixed $cz$, interpolating between the early onset of chaos and late time behavior, (3) we explicitly compute the Virasoro vacuum block to order $1/c^2$ and $1/c^3$, corresponding to $2$ and $3$ loop calculations in AdS$_3$,  and (4) we derive the   heavy-light vacuum blocks in theories with $\mathcal{N}=1,2$ superconformal symmetry.
}

\arxivnumber{}

\begin{document}

\maketitle
\flushbottom
 
\section{Introduction and Summary}

The infinite dimensional Virasoro algebra profoundly contrains the dynamics of Conformal Field Theories (CFTs) in two dimensions.  Certain ``rational'' theories have operator algebras that truncate, allowing them to be solved exactly.   But despite their phenomenological relevance and beauty, rational theories are small islands in a largely uncharted sea of 2d CFTs.   Furthermore, we can only study quantum gravity in AdS$_3$ by analyzing CFTs with large central charge $c$, and relatively little is known about these `irrational' 2d CFTs.
%
%

Although it appears that large $c$ CFTs cannot be solved exactly,   it is still possible to take some of the methods \cite{Belavin:1984vu,Zamolodchikov:1987}  that make rational CFTs tractable and apply them \cite{Fitzpatrick:2016ive} to irrational theories. The reason is that correlation functions in any CFT$_2$ can be decomposed into Virasoro conformal blocks $\CV_{h_i,h,c}(z)$ as\footnote{We have explicitly indicated the decomposition into a product of holomorphic and anti-holomorphic parts.}   
\be \label{eq:GeneralBlockDefinition}
\< \CO_1(\infty) \CO_2(1) \CO_3(z) \CO_4(0) \> = \sum_{h, \bar h} P_{h, \bar h} \mathcal{V}_{h_i, h, c}(z)  \mathcal{V}_{\bar h_i, \bar h, c}(\bar z) .
\ee
The Virasoro blocks are the contributions to the Operator Product Expansion (OPE) of $\CO_3(z) \CO_4(0)$  from irreducible representations of the Virasoro algebra
 \be 
 [L_n, L_m] &=& (n-m) L_{n+m} + \frac{c}{12} n (n^2-1) \delta_{n+m,0}
 \label{eq:VirasoroAlgebra}
 \ee
The $h_i, \bar{h}_i$ are weights of the external operators $\CO_i$, while $h, \bar h$ are intermediate operator weights. When $\CO_1=\CO_2$ and $\CO_3=\CO_4$, a universal contribution in equation (\ref{eq:GeneralBlockDefinition}) is the Virasoro vacuum block, which  encapsulates the exchange of any number of pure AdS$_3$ `graviton' states between the external operators. 

The Virasoro blocks have turned out to be extremely useful as a source of information about gravity in AdS$_3$, and in fact BTZ black hole \cite{BTZ} thermodynamics \cite{Hartman:2014oaa} emerges in a universal, theory-independent way from the heavy-light, large central charge limit of the Virasoro blocks \cite{Fitzpatrick:2014vua, Fitzpatrick:2015zha, Fitzpatrick:2015foa, Alkalaev:2015wia, Alkalaev:2015lca, Fitzpatrick:2015dlt, KrausBlocks, Beccaria, Anous:2016kss, Besken:2016ooo}.  Information loss from black hole physics appears to be due to the behavior of the blocks in this limit \cite{Fitzpatrick:2015zha, Fitzpatrick:2015dlt, Fitzpatrick:2016ive}.   The blocks are also the basic components of the conformal bootstrap program \cite{FerraraOriginalBootstrap1, PolyakovOriginalBootstrap2, Rattazzi:2008pe, Rychkov:2016iqz, Simmons-Duffin:2016gjk}.  Knowing their explicit forms would greatly assist the study of 2d CFTs and 3d gravity using the bootstrap \cite{Hellerman:2009bu, Hartman:2014oaa, Jackson:2014nla, Chang:2015qfa, Lin:2015wcg, Benjamin:2016fhe, Chang:2016ftb}.

Each conformal block depends only on the quantum numbers $(h_i,h,c)$ of the representations involved and not on the specific theory. 
A  strategy for computing the blocks is to work with a theory where operator truncation occurs, and then use the fact that the result is theory-independent.  This technique becomes even more useful when augmented by the fact that the conformal blocks are analytic functions of their defining quantum numbers, so that one can compute the blocks for special values of the external dimensions $h_i$ and then analytically continue. In this paper, we will use this technique to develop an efficient method to compute and study the blocks order-by-order in a $1/c$ expansion, and to perform certain all-orders Lorentzian resummations.\footnote{Currently, closed form expressions for Virasoro blocks have been obtained in an expansion about various limits, such as  $h \to \infty$ \cite{Zamolodchikovq}, as well as at $c \to \infty$ in the all light and the heavy-light limit \cite{Fitzpatrick:2014vua, Fitzpatrick:2015zha, Fitzpatrick:2015foa, Alkalaev:2015wia, Alkalaev:2015lca, KrausBlocks, Beccaria,Fitzpatrick:2015dlt}. In addition, as a function of $c$ and of intermediate operator dimensions, the Virasoro blocks are meromorphic functions with only simple poles. 
These properties imply recursion relations \cite{ZamolodchikovRecursion, Zamolodchikovq} that efficiently compute the series expansion \cite{Perlmutter:2015iya} of the vacuum blocks near $z=0$ with generic $h_i, h, c$.
}

We will organize the series expansion in terms of the ansatz\footnote{In this paper, we denote by $\cal V$ the vacuum Virasoro block component of the correlator $\frac{\left\langle\CO_L(0)\CO_L(z)\CO_H(1)\CO_H(\infty)\right\rangle}{\left\langle\CO_H(1)\CO_H(\infty)\right\rangle}$, while we use $\tilde{\cal V}$ for the normalized vacuum block, i.e. the vacuum block component of $\frac{\left\langle\CO_L(0)\CO_L(z)\CO_H(1)\CO_H(\infty)\right\rangle}{\left\langle\CO_L(0)\CO_L(z)\right\rangle\left\langle\CO_H(1)\CO_H(\infty)\right\rangle} \supset \tilde{\cal V}$ begins with $\tilde{\cal V} = 1 + \cdots$ in the small $z$ expansion.} 
\begin{align}\label{eq: ansatzV}
\mathcal{V}_{h_H,h_L,0,c}(z) & =\exp\left[h_L \sum_{n,m=0}^{\infty}\left(\frac{1}{c}\right)^{m}\left(\frac{h_{L}}{c}\right)^{n}f_{mn}\left(\eta_H,z\right)\right],
\end{align}
where $\eta_H=\frac{h_H}{c}$ is fixed at large $c$, and we will compute the function $f_{mn}$. 
This is a natural expansion to use in the semiclassical limit \cite{Zamolodchikovq,HartmanLargeC,HarlowLiouville,Fitzpatrick:2014vua,Beccaria}, which keeps only the terms with $m=0$, but direct calculations \cite{Fitzpatrick:2015foa} indicate that it is also justified to all orders in the $1/c$ expansion (see section \ref{sec:DegenerateOperators} for a more detailed discussion).  We explicitly compute up to order $1/c^3$, i.e. $m+n \le 3$, relegating many of the detailed forms of the functions $f_{mn}$ to appendix \ref{appendixfmnk}.  

We have verified that our results match with a direct computation of the blocks to high orders in a series expansion in $z$, providing further direct evidence for the validity of our methods and for their application to information loss \cite{Fitzpatrick:2016ive}.  We also apply this method to compute the heavy-light super-Virasoro vacuum block in the case of $\CN=1$ and $\CN=2$ superconformal symmetry.  With some straightforward but tedious work, the method could certainly be extended to theories with more supersymmetry, which have been studied recently using the bootstrap \cite{Lin:2015wcg}.

One physically interesting regime  where our techniques prove to be particularly effective is in the limit where $z \to 0$  in the Lorentzian region.  This limit would be trivial in the Euclidean regime, $\bar{z} = z^*$, where $z\rightarrow 0$ is the OPE limit of a conformal block and is therefore dominated by the primary state contribution. However, the regime of small $z$ becomes highly non-trivial after analytically continuing through a branch cut to the Lorentzian sheet.  In particular, correlators in the Lorentzian regime depend on the order of the operators, and continuing along different paths before taking $z\rightarrow 0$ can produce different results, which generally include singularities at small $z$.  


This behavior is related to a variety of fascinating physical phenomena, including bulk singularities \cite{GGP}, black hole scrambling \cite{Roberts:2014ifa}, and universal CFT causality constraints \cite{Hartman:2015lfa}, to name a few.  This regime was  studied at subleading order in the large $c$, heavy-light limit of the vacuum block \cite{Fitzpatrick:2016thx}, where it was found that certain $\frac{1}{c} \log(z)$ terms appear.  These  were argued to be $1/c$ corrections to the power-law behavior of  singular terms.  In particular, at leading non-trivial order the growth of the singular terms is $\sim z^{-1}$, which after mapping to the thermal cylinder $z = e^{ 2 \pi i (t+x)/\beta}$ corresponds to  exponential growth with ``Lyapunov'' exponent $\frac{2 \pi}{ \beta}$.  In \cite{Fitzpatrick:2016thx} a  logarithmic correction at the next order in $1/c$ was argued to be the leading term in a correction to this exponent, shifting it to $\frac{2\pi}{\beta} ( 1+ \frac{12}{c})$.  In section \ref{sec:AllOrdersResummation}, we will prove that there are indeed an infinite series of terms of the form 
\be
\frac{1}{c z} \left(\frac{\log(z)}{c} \right)^n
\ee with exactly the correct coefficients to resum into a correction to the Lyapunov exponent.  This might also be viewed as a quantum correction to the Regge trajectory.  We will also provide a simple way to understand subleading logarithms, ie terms  of the form $\frac{1}{c^m} \left( \frac{\log(z)}{c} \right)^n$ with $m>1$.  

As noted in \cite{Fitzpatrick:2016thx}, there are also power-law corrections that are larger than the logarithmic corrections.  In the limit $c \rightarrow \infty$ with $c z$ fixed, there is an infinite sequence of terms of the form $(c z)^{-n}$ that survive.  The ``Lyapunov'' regime, where the onset of scrambling first takes place, is the regime of large $c z$ and is well-described by the first few terms in a $1/c$ expansion.  However, eventually the behavior transitions to the ``Ruelle'' regime \cite{JoeRuelle}, related to the decay of quasi-normal mode excitations around a BTZ black hole, and to describe this regime of small $c z$ one must sum all the leading terms.  As we will see, this series is asymptotic, so one must Borel resum it.  We will show how to do this in subsection \ref{sec:leadingsing}, with a remarkably simple result that interpolates between the ``Lyapunov'' regime and the ``Ruelle'' regime:
\be
\lim_{c \rightarrow \infty \atop c z \textrm{ fixed} } z^{2h_L} \CV_{h_H,h_L,0}(z) &=&  G(h_H, h_L, \frac{i c z}{12\pi} ) + G(h_L,h_H, \frac{i c z}{12\pi}), \nn\\
G(h_1,h_2,x) &\equiv & (x )^{2h_1} (2h_2)_{-2h_1} ~ {}_1F_1(2h_1,1+2h_1-2h_2, x).
\ee
It should be remembered, however, that non-vacuum blocks may also significantly affect the behavior of the correlator at intermediate and late times.

The idea that makes these Lorentzian resummations possible is that analytic continuation from the Euclidean to the Lorentzian region simply transforms a degenerate vacuum block into a finite sum of degenerate blocks.  In other words, when evaluated on the second (Lorentzian) sheet, the degenerate vacuum block $\CV_{(1,s)}(z)$ becomes a linear combination of  $s$ degenerate blocks, which are to be evaluated on the first (Euclidean) sheet.  Once we understand the behavior of $\CV_{(1,s)}(z)$ in the Lorentzian region for all $s$, we can use this to obtain the physical vacuum blocks with general $h_L$.  We justify and implement these ideas in section \ref{sec:AllOrdersResummation}.

The outline of this paper is as follows.  In section \ref{sec:DegenerateOperators} we review degenerate operators and outline our method of computation.  The in section \ref{sec:ComputingExpansion} we use the method to compute the heavy-light vacuum Virasoro block at order $1/c$, and the all-light Virasoro block up to order $1/c^3$, which would correspond to a 3-loop gravitational calculation in AdS$_3$.  We use two methods, one based on solving differential equations, and another based on a $1/c$ expansion of the Coulomb gas formalism.  In section \ref{sec:AllOrdersResummation} we state and prove various results on the resummation of logarithms and singularities in the Lorentzian regime, and discuss the application of these results to the study of quantum chaos.  Finally, in section \ref{sec:SuperVirasoro} we derive the super-Virasoro vacuum block for $\CN =1, 2$ superconformal symmetry.  Various technical details have been relegated to the appendices.

\section{Degenerate Operators and Heavy-Light Virasoro Blocks}
\label{sec:DegenerateOperators}

In this section, we will review the properties of degenerate operators\footnote{See \cite{DiFrancesco:1997nk} or \cite{Ginsparg} for more systematic reviews.} and explain how to use them to extract information concerning the Virasoro vacuum block in the large central charge or $c \gg 1$ limit.  A degenerate operator is a Virasoro primary operator with null descendants, which means that some of its Virasoro descendants have vanishing norm. When discussing degenerate states it is useful to introduce a parameter $b$ so that
\begin{equation}
c \equiv 1 + 6 \left(  b + \frac{1}{b} \right)^2 .
\end{equation}
In this work, we take the $c \to \infty$ limit via $b \to \infty$.  In this notation, the simplest example of a null state is the second level descendant
\be
\label{eq:12NullDescendant} 
\left( L_{-1}^2 + b^2 L_{-2} \right) | h_{1,2} \> = 0 .
\ee
One can check using the Virasoro algebra of equation (\ref{eq:VirasoroAlgebra}) that the level 2 Gram matrix
\be
\label{eq:Level2Det} 
\left( \begin{array}{cc}
\< h | L_1^2 L_{-1}^2 | h\>  & \< h | L_1^2 L_{-2} | h\>  \\
\< h | L_2 L_{-1}^2 | h\> & \< h | L_2 L_{-2} | h\>  \end{array} \right)
\ee
has a vanishing determinant when the holomorphic dimension satisfies $h_{1,2} = -\frac{1}{2} - \frac{3}{4b^2}$; the level two descendant in equation (\ref{eq:12NullDescendant}) is the corresponding null vector.  
In general, degenerate states  can only occur for holomorphic dimensions satisfying the Kac formula
\be
\label{eq:GeneralDegenerateDimension}
h_{r,s} &=& \frac{b^2}{4}(1-r^2)  + \frac{1}{4b^2} (1-s^2) + \frac{1}{2}(1-rs)
\ee
for positive integers $r,s$.  This formula determines the values of dimension $h$ when the Kac determinant, of which equation (\ref{eq:Level2Det}) is an elementary example, vanishes.  Notice that  $r \leftrightarrow s$ simply corresponds with $b \leftrightarrow 1/b$. For rational models, $b^2 (\equiv - \frac{p}{p'})$ is a rational number, and consequently so are $h_{r,s}$ and $c$.  In this work we will mainly be interested in general (irrational) values of $b$ and $h_{r,s}$.

Null conditions such as (\ref{eq:12NullDescendant}) translate into differential equations for the correlation functions involving a degenerate state.  This follows because within a correlator with operators of dimension $h_i$, the Virasoro generators $L_{-m}$  act as differential operators due to the stress tensor Ward identities.
 In the simplest case of $\mathcal{O}_{1,2}$, we have: 
\begin{equation}
\label{eq:NullDiffEqh21}
\left( \partial_z^2 + \left(2 \frac{1+b^{-2}}{z} + \frac{b^{-2}}{1-z}\right) \partial_z  + \frac{b^{-2} h_H}{(1-z)^2} \right) \frac{\< \CO_H(\infty) \CO_H(1) \CO_{1,2}(z) \CO_{1,2}(0) \>}{\< \CO_H(\infty) \CO_H(1) \>\< \CO_{1,2}(z) \CO_{1,2}(0) \>} = 0.
\end{equation}
At $b\rightarrow\infty$, $\CO_{1,2}$ has dimension $h_{1,2}\rightarrow -\frac{1}{2}$ and is a light ``probe'' operator. The other operator, $\CO_H$, has arbitrary weight $h_H$.  Equation (\ref{eq:NullDiffEqh21}) is a version of the hypergeometric differential equation; it is an exact relation for this correlator and its conformal blocks.  One of its solutions, corresponding to the vacuum conformal block, is given by
\begin{equation}
\label{eq:IntroExample}
\frac{\< \CO_H(\infty) \CO_H(1) \CO_{1,2}(z) \CO_{1,2}(0) \>}{\< \CO_H(\infty) \CO_H(1) \>\< \CO_{1,2}(z) \CO_{1,2}(0) \>} = (1-z)^{\frac{\beta_H}{b}} {}_2F_1\left( 1+b^{-2}, 2\beta_H, 2 (1+ b^{-2}), z\right),
\end{equation}
where $\beta_H$ is a parameterization of the operator dimension $h_H$ and is related to its Coulomb gas charge:
\be\label{eq:alphaH}
\beta_H &=& \frac{1}{2b} \left( Q - \sqrt{Q^2-4h_H} \right), \nn\\
h_H &=& b \,\beta_H( Q-b \,\beta_H), \nn\\
 Q&\equiv& b + b^{-1} .
\ee  
 
We will be interested both in the ``light-light'' limit, where $h_H $ is $\CO(1)$, as well as in the ``heavy-light'' limit where $b^{-2} h_H$ is fixed in the large $b$ limit. In the heavy-light limit, $\CO_H$ represents a heavy operator generating a background probed by $\CO_{1,2}$.  More specifically, in a putative AdS$_3$ dual description, $\CO_H$ will create either a deficit angle or BTZ black hole \cite{BTZ}.   At $c\rightarrow \infty$ in the heavy-light limit, (\ref{eq:IntroExample}) simplifies to
 \be
e^{-\frac{1}{2} t_E} \frac{\sin(\pi T_H t_E)}{\pi T_H} ,
\ee
where $t_E=-\log(1-z)$ is the Euclidean time and $
T_H=\frac{1}{2\pi}\sqrt{\frac{24h_H}{c}-1}$ is the Hawking temperature of a BTZ black hole created by acting with $\CO_H$ on the vacuum. 

More generally, the vacuum block for the correlator $\< \CO_H(\infty) \CO_H(1) \CO_{r,s}(z) \CO_{r,s}(0)\>$ satisfies a finite order differential equation for all of the degenerate operators $\CO_{r,s}$.  Since the conformal blocks depend only on the parameters $h_i, h_p, b$, and not on the particular theory, this suggests that one can compute them in general by solving the resulting differential equations.  Of course, there is an obvious obstacle: the light weights $h_{r,s}$ are not quite independent free parameters.  We can dial their value by changing their indices $r$ and $s$, but within some limitations.  First, $r$ and $s$ must be positive integers, and at large $c>0$ this means $h_{r,s}$ are always in the non-unitary regime.  This is not as significant a limitation as it may seem, because the conformal blocks are meromorphic functions of $h_{r,s}$ (for a detailed discussion see \cite{Fitzpatrick:2016ive}).  Thus, one can hope to analytically continue the blocks as a function of integer $(r,s)$ to non-integer values.\footnote{Continuing a function on the integers to the entire complex plane generally requires some additional knowledge of its behavior at $\infty$; we will see that order-by-order in the large $c$ expansions we employ in this paper, the required information is provided by the OPE.} 
And in fact this method was used in \cite{Fitzpatrick:2016ive} to study contributions to the vacuum block that are non-perturbatively small in the large $c$ limit, which are associated with the resolution of information loss problems.  

A second, more serious obstacle is that increasing $r$ and $s$ produces new differential equations of increasingly high orders.  Thus, solving for more values of $h_{r,s}$ requires solving increasingly complicated differential equations of increasingly high order.  We will see that this translates into increasing complexity in using the method to solve for the vacuum block at increasingly high orders in $1/c$.  Nevertheless, comparison with other methods \cite{Fitzpatrick:2014vua, Fitzpatrick:2015zha, Fitzpatrick:2015foa} suggests that this may be the most efficient available procedure for determing the large $c$ vacuum blocks, especially if one wishes to go beyond the semi-classical limit.


To be more precise about the method we use, we write a generic vacuum block (which will not involve any degenerate operators) in a double expansion in $\frac{1}{c}$ and $\frac{h_L}{c}$:\footnote{A limitation of this method is that it is much more complicated to get results for general internal dimension $h_I$ of the conformal block.  The reason for this is that once the external dimensions $h_L, h_H$ are fixed and $\CO_L$ is chosen to be a degenerate operator, then the dimensions of the allowed internal operators are also fixed to lie in a finite set.  In principle, one could hope to get around this by using the fact that degenerate operators $\CO_{1,s}$ contain more and more operators in the OPE as $s$ is increased, and in the limit that $s$ becomes large one would have access to a tower of operators with a discretum of dimensions.  However, each order in $1/c$ has a complicated dependence on $h_I$, in contrast to the simple polynomial dependence on the external dimensions $h_L, h_H$ (for instance,  the $c\rightarrow \infty$ piece is the global block, which is independent of $h_H$ and $h_L$ but a hypergeometric function ${}_2F_1(h_I, h_I, 2h_I,z)$ of $h_I$), so extracting this dependence from the discretum of exchanged operators really requires the entire infinite tower, which in turn requires solving the large $c$ degenerate blocks in the $s\rightarrow \infty$ limit.  Thus we are focusing entirely on the vacuum Virasoro block in this paper.} 
\begin{align}\label{eq:LargeCAnsatz}
\mathcal{V}_{h_H,h_L,0,c}(z) & =\exp\left[h_L \sum_{n,m=0}^{\infty}\left(\frac{1}{c}\right)^{m}\left(\frac{h_{L}}{c}\right)^{n}f_{mn}\left(\eta_H,z\right)\right],
\end{align}
where $\eta_H=\frac{h_{H}}{c}$.  This ansatz can be justified as follows.  In the semiclassical limit of $c \to \infty$ with all $h_i/c$ fixed, we have a great deal of evidence \cite{Zamolodchikovq,HartmanLargeC,HarlowLiouville,Fitzpatrick:2014vua,Beccaria} that the vacuum block can be written as $\exp \left( c \, g(\frac{h_L}{c}, \frac{h_H}{c}, z ) \right)$ for some function $g$ that is analytic in $h_L/c$ and $h_H/c$ in a neighborhood around the origin.  This explains why equation (\ref{eq:LargeCAnsatz}) does not contain terms such as e.g. $h_L^4/c^2$, which would behave very differently in the semi-classical limit.  Corrections to the semi-classical limit can then be expanded in powers of $1/c$, leading to equation (\ref{eq:LargeCAnsatz}).  Note that the exponential form of the ansatz is convenient, but beyond the semiclassical limit it is not obligatory.  One can also justify the ansatz to any order in $z$ via a direct, brute force computation \cite{Fitzpatrick:2015foa} of the vacuum block using the Virasoro algebra.  Finally, note that although we have expanded the ansatz in a heavy-light limit, the conformal blocks will be symmetric under $h_L \leftrightarrow h_H$.

As mentioned previously, the vacuum block in equation (\ref{eq:LargeCAnsatz}) is analytic in $h_L$ and $h_H$ \cite{Zamolodchikovq, Fitzpatrick:2016ive}.   Therefore, when taking $h_L=h_{r,s}$, we must recover the null state vacuum block such as solution (\ref{eq:IntroExample}). Matching order by order in $\frac{1}{c}$ and $\frac{h_L}{c}$ to these solutions, we can determine $f_{mn}(\eta_H, z)$.  Note that our knowledge of the block in the heavy-light semiclassical limit strongly constrains its behavior at large values of the external dimensions, so it seems very unlikely that there are any ambiguities in the analytic continuation from $h_L = h_{r,s}$.

The method can be generalized to study theories with supersymmetry. In particular, we work out heavy-light large $c$ limit of the holomorphic part of super-Virasoro vacuum blocks with $\mathcal{N}=1,2$ supersymmetries in section \ref{sec:SuperVirasoro}.  It turns out that the super-Virasoro vacuum block of the lowest component fields in these theories do not get contributions from the fermionic supersymmetry generators at leading order of the large $c$ limit, so they largely match with results extrapolated from  \cite{Fitzpatrick:2015zha}, but it is interesting to understand the supermultiplet structure and the correlators of superconformal descendant fields. 

Although the method is straightforward, it becomes quite tedious beyond the first few orders in (\ref{eq:LargeCAnsatz}). When $r$ is large, it becomes a non-trivial task to construct the null state differential equation for $\phi_{1,r}$, which is a complicated $r$-th order differential equation whose exact solutions can be difficult to compute. But in specific limits of physical interest, these equations simplify greatly and become extremely useful in determining key properties of the higher order quantum corrections. One example of these are the leading $\log$ terms in the $\frac{1}{c}$ corrections when all four operators are light, which we discuss in section \ref{sec:AllOrdersResummation}. Such terms plays an important role in the growth of quantum chaos \cite{Maldacena:2015waa, Roberts:2014ifa, Fitzpatrick:2016thx} and can be computed efficiently with the $\phi_{1,r}$ null state differential equations. 

Another very useful way to get higher order corrections in the large $c$ limit is to use the Coulomb gas formalism, which provides a straight-forward construction of integral representations for the Virasoro blocks involving degenerate operators. We have used it in \cite{Fitzpatrick:2016ive} in order to study the non-perturbative part of the vacuum Virasoro block in the large $c$ asymptotic expansion. In this work, we will show that directly expanding the integrand in the Coulomb gas formalism provides an efficient way to obtain higher order terms in (\ref{eq:LargeCAnsatz}).  This method is discussed in section \ref{sec:CoulombGasMethod}.

\section{Computing the $1/c$ Expansion of the Vacuum Block}
\label{sec:ComputingExpansion}

In this section, we will use the computational method explained in last section to calculate the higher order corrections to the Virasoro block.
The idea is to assume that the general heavy-light vacuum block $\mathcal{V}$
can be written as the ansataz (\ref{eq:LargeCAnsatz}). When $\mathcal{O}_{L}$ is a degenerate operator, $\mathcal{V}$ satisfies a null-state differential equation. At order $\frac{1}{c^{p}}$,
there are $p+1$ $f_{mn}$ functions $\{f_{0,p}, f_{1,p-1}, \dots, f_{p,0}\}$. Each one appears with a different power of $h_L$ in its coefficient, i.e. $\log \CV \supset h_L^{n+1} f_{m,n}$.  By (\ref{eq:GeneralDegenerateDimension}), the degenerate operators $h_{1,s}$ with $r=1$  have weights
\be
h_{1,s} = \frac{1}{2}(1-s)+ \frac{1}{4b^2} (1-s^2) \approx \frac{1}{2} (1-s) + \frac{3}{2c} (1-s^2) + \CO\left(\frac{1}{c^2}\right),
\ee
that are $\CO(1)$ at  $c\rightarrow \infty$.  For any choice of $s$, the operator $\CO_{1,s}$ produces a differential equation that we can solve for $\CV$ and expand at large $c$ to obtain the $\CO(c^{-p})$ term as
\be
\label{eq:CVnewterm}
\log \CV \supset \frac{1}{c^p} \left( h_{1,s} f_{p,0}+ h_{1,s}^2 f_{p-1,1} + \dots + h_{1,s}^{p+1} f_{0,p}\right),
\ee  
Unfortunately, for a single fixed $s$, $h_{1,s}$ is just a number and therefore knowledge of (\ref{eq:CVnewterm}) does not allow one to separate out the different contributions $f_{mn}$.\footnote{More precisely, $h_{1,s}$ is not just a fixed number but rather a fixed function of $c$.  However, because of the relation
\be
h_{1,s} &=& h_{1,s}^{(0)} + \frac{1}{b^2} (h_{1,s}^{(0)} - (h_{1,s}^{(0)})^2),
\ee
where $h_{1,s}^{(0)} = \lim_{c\rightarrow \infty} h_{1,s}$, we are free to perform an expansion in powers of $h_{1,s}$ or in powers of $h_{1,s}^{(0)}$, since the difference between the two just corresponds to a redefinition of the $f_{m,n}$ functions.} 
  To accomplish this, one needs to take $p+1$ different degenerate operators, which give $p+1$ differential equations to be solved for these $p+1$ $f_{mn}$ functions.   
  
  This is the procedure that we will implement in sections \ref{sec:VacBlockFirstOrderDiff} and \ref{sec:VacBlockThirdOrderDiff} in order to obtain $1/c^p$ corrections up to $p=3$, corresponding to 3-loop gravitational effects in AdS$_3$.  In section \ref{sec:CoulombGasMethod} we will study a different method that uses the Coulomb gas formalism to replace differential equations with integrals.

A convenient and efficient formalism for keeping track of the null state differential equations at large $c$ was developed in \cite{Bauer:1991ai, DiFrancesco:1997nk}.  Let $D_{1,s}$ be the following matrix:
\be 
D_{1,s} = -J_- + \sum_{m=0}^\infty\left(\frac {J_+}{b^2}\right)^m L_{-m-1},  
\ee
where $J_\pm$ are matrix generators of the spin $(s-1)/2$ representation of $SU(2)$:
\be
(J_0)_{ij} &=& \frac{1}{2} (s-2i+ 1)\delta_{ij}, \nn\\
(J_-)_{ij} &=& \left\{ \begin{array}{cc}  \delta_{i,j+1} & (j=1,2, \dots, s-1) \\ 0 & \textrm{else}\end{array} \right. ,\qquad \qquad  \qquad \qquad \begin{array}{c} \left[J_+, J_- \right] = 2 J_0 , \\ \left[J_0, J_\pm \right] = \pm J_\pm . \end{array}  \nn\\
(J_+)_{ij} &=& \left\{ \begin{array}{cc} i(s-i)\delta_{i+1,j} & (i=1, 2, \dots, s-1), \\ 0 & \textrm{else} \end{array} \right.  . 
\ee
Then, the null state equation of motion is given by the equation $f_0 = 0$ after eliminating $f_1, \dots, f_{s-1}$ from the equation
\be
D_{1,s} \left( \begin{array}{c} f_1 \\ f_2 \\ \vdots \\ f_s \end{array} \right) &=& \left( \begin{array}{c} f_0 \\ 0 \\ \vdots \\ 0 \end{array} \right) .
\ee

At infinite $c$ with $h_H$ held fixed and $\CO(1)$, one can manifestly drop all terms in the sum in $D_{1,s}$ except for $m=0$, so the null state manifestly becomes
\be
\label{eq:largecnull1}
L_{-1}^s \CO_{1,s}=0,
\ee
and the infinite $c$  differential equation for the conformal block becomes
\be
\label{eq:largecnull2}  
\partial_z^s  \CV(z) =0,
\ee
where the factor of $z^{s-1}$ arises because our convention for $\CV(z)$ factors out the $\<\CO_{1,s}(z) \CO_{1,s}(0)\>$ two-point function.  
More generally, allowing $h_H$ to be $\CO(c)$ with $h_H/c$ fixed, at infinite $c$ the differential equation for $\CV(z)$ becomes \cite{Fitzpatrick:2016ive,Bauer:1991ai, DiFrancesco:1997nk}
\be
\label{eq:leadinglight}
\left[ \prod_{k=-(s-1) + 2j \atop j=0, \dots , s-1} \left( \partial_t - \frac{k}{2} \sqrt{ 1-24\eta_H} \right) \right] e^{\frac{s-1}{2} t} \CV(t) =0,
\ee
where $t=-\log(1-z)$. 
 
Let us see how this works in the simplest case, namely at lowest order in $1/c$ in (\ref{eq:LargeCAnsatz}).  At order $c^{0}$, $\text{log}\CV\supset h_L f_{00} \left(\eta_H,z\right)$, there is only one unknown function $f_{00}$, which means that we only need the differential
equation (\ref{eq:leadinglight}) with $s=2$:\footnote{
For later reference, the exact equation for the vacuum block for $\CO_{1,2}$ is
\begin{equation}
\left( \partial_z^2 + \left(2 \frac{1+b^{-2}}{z} + \frac{b^{-2}}{1-z}\right) \partial_z  + \frac{b^{-2} h_H}{(1-z)^2} \right)\left(z^{2h_{1,2}}\CV(h_{1,2},\eta_H,z)\right)=0.
\label{eq:null12}
\end{equation}
with $b^2=-\frac{3}{2 (2h_{1,2}+1)}$ in this equation.
}
\begin{equation}
\left( \frac{d^2}{dt^2} - \frac{1-24 \eta_H}{4}\right) e^{\frac{t}{2}} e^{h_{1,2} f_{00}(z(t))},
\end{equation}
with $h_{1,2}\simeq-\frac{1}{2}$. 
In terms of $z= 1- e^{-t}$, we obtain
\begin{equation}
f_{00}''=\frac{12\eta_H}{\left(1-z\right)^{2}}+\frac{1}{2}\left(f_{00}'\right)^{2}.\label{eq:nullf00}
\end{equation}
And the solution that corresponds to the vacuum block is 
\begin{equation}
f_{00}\left(\eta_H,z\right)=-\left(1-2\pi i T_H \right)\log\left(1-z\right)-2\log\left(\frac{1-\left(1-z\right)^{2 \pi i T_H}}{2 \pi i T_H}\right). \label{eq:f00}
\end{equation}
with $T_H=\frac{1}{2\pi} \sqrt{24\eta_H-1}.$ This reproduces the result for the heavy-light limit of the vacuum block first found in \cite{Fitzpatrick:2014vua}.

\subsection{The Heavy-Light Virasoro Vacuum Block at Order $1/c$}
\label{sec:VacBlockFirstOrderDiff}

At order $1/c$ , there are two functions $f_{01}$ and $f_{10}$:
\[
\text{log}\CV\supset \frac{h_L}{c}\left(f_{10}+h_Lf_{01}\right)
\]
which means that we need to use both the $h_{1,2}$ null-state equation  (\ref{eq:null12}) and the $h_{1,3}$ null-state equation:
\be
0 &=& \left( \frac{1}{z^2} \frac{d^3}{dz^3} z^2  + \left( \frac{ 4 \frac{1}{b^2}  \frac{2-z}{1-z}}{z} \right) \frac{d^2}{dz^2} + \frac{4 (\frac{1}{b^2} + \frac{2}{b^4} ) h_H (2-z)}{(1-z)^3 z} \right. \nn\\
&&  \left. + \left( \frac{2 \frac{1}{b^2} (9-13z+(5+2h_H)z^2)+ \frac{1}{b^4}(3+(z-3)z)}{z^2 (1-z)^2} \right) \frac{d}{dz}  \right)  \left(z^{2h_{1,3}}\CV\left({h_{1,3},\eta,z}\right)\right)\nn\\
\label{eq:null13}
\ee 
with 
$b^2=-\frac{2}{h_{1,3}+1}$ in this equation and $h_{1,3}\simeq-1-\frac{12}{c}-\frac{156}{c^{2}}+\CO\left(1/c^{3}\right)$ in the large $c$ limit.
The $c^{0}$ order of equation (\ref{eq:null13}) only involves $f_{00}$ and the solution for it is exactly equation (\ref{eq:nullf00}). 

At order $1/c$, equation (\ref{eq:null12}) and (\ref{eq:null13}) give the following two equations for $f_{00},f_{01}$ and $f_{10}:$ 
\begin{equation}
\begin{aligned}
0&=F_1''-f_{00}' F_1'-8 f_{00}''-5 f_{00}'{}^2-\frac{12(2z-1)}{z(z-1)} f_{00}'-\frac{12 (z+1)}{(z-1) z^2}\\
0&=F_{2}'''-3f_{00}'F_{2}''+3\left(f_{00}'^2-f_{00}''+\frac{8 \eta_H }{(z-1)^2}\right)F_{2}'+12f_{00}'''+\left(\frac{24-48 z}{(z-1) z}-72 f_{00}'\right) f_{00}''+36f_{00}'{}^3\\
&\quad +\frac{24(2z-1)}{z(z-1)}f_{00}'{}^2+\frac{12 ((50 \eta_H +3) z^2-z-1)}{(z-1)^2 z^2} f_{00}'+\frac{24 \left((25 \eta_H +1) z^3-2 z+1\right)}{(z-1)^3 z^3}\label{eq:null13c1}
\end{aligned}
\end{equation}
where we define
\[
F_{1}=2f_{10}-f_{01},F_{2}=f_{10}-f_{01}.
\]
Note that the differential equations \eqref{eq:null13c1} involve the `zeroth order' term $f_{00}$, which also appears at higher orders, since $\log\CV\supset h_{1,s}f_{00}=(\frac{1-s}{2}+\frac{1-s^2}{4b^2})f_{00}$. There are a few significant simplifications that occurred in the above equations.  First,  $f_{10}$ and $f_{01}$  show up only as a certain combination
($F_{1}$ and $F_{2}$) in each equation. The reason is that these
equations come from the leading term in $\frac{h_{1,s}f_{10}}{c}+\frac{h_{1,s}^{2}f_{01}}{c}$.
Since $h_{1,s}=\frac{1-s}{2}+\CO\left(1/c\right)$, the leading
term in $\frac{h_{1,s}f_{10}}{c}+\frac{h_{1,s}^{2}f_{01}}{c}$is 
\[
\frac{\frac{1-s}{2}f_{10}}{c}+\frac{\left(\frac{1-s}{2}\right)^{2}f_{01}}{c}.
\]
A similar phenomenon continues to be true for higher order calculations.
This means that these differential equations can be solved independently for $F_1$ and $F_2$. Second, 
 only the derivatives of $F_{1}$ and $F_{2}$ show up in these
equations.  This allows one to solve for the derivatives first, and then integrate.  We have found this allows one to obtain a closed form expression for $F_1(z)$ directly using Mathematica; on the other hand, the differential equation for $F_2$ is too complicated to be solved this way.  Since the solutions are known from previous work \cite{Fitzpatrick:2015dlt} (see also \cite{Beccaria,Hijano:2015rla} for semi-classical results), one can substitute them into equations (\ref{eq:null13c1}) and verify them.  For completeness, these solutions are included in appendix \ref{appendixfmnk}.

\subsection{The All-Light Virasoro Vacuum Block at Order $1/c^2$ and $1/c^3$}
\label{sec:VacBlockThirdOrderDiff}


At order $1/c^2$, there are three functions $f_{20},f_{11}$ and $f_{02}$:
\[
\log\CV\supset \frac{h_L}{c^2}(f_{20}+h_L f_{11}+h_L^2 f_{02}).
\]
To fully determine them, one needs to solve the $h_{1,2}$ and $h_{1,3}$ null-state equations and also the $h_{1,4}$ null state equation at order $1/c^2$. These equations are complicated, but at least one can expand them in terms of $\eta_H \equiv \frac{h_H}{c}<1$ and obtain the result as an expansion in $\eta_H$.  
Define the expansion of $f_{mn}$ as 
\begin{equation}
\begin{aligned}
f_{mn}=&\sum_{k=0}^{\infty}\eta_H^{k+1}f_{mnk}\qquad \text{for}\ m \text{ or } n>0\\
f_{00}=&-2\log(z)+\sum_{k=0}^{\infty}\eta_H^{k+1}f_{00k}
\end{aligned}
\end{equation}
where the $-2\log(z)$ in $f_{00}$ is because we include the prefactor $z^{-2h_L}$ in the definition of the vacuum block. Since the vacuum block $\CV\left(h_{L},h_{H},z\right)$ is symmetric under the exchange $h_{L}\leftrightarrow h_{H}$, in our convention, this means that $f_{ijk}=f_{ikj}$.

The liner $\eta_H$ and $\eta_H^2$ terms at order $1/c^2$ are 
\[
\text{log}\CV\supset \frac{h_L}{c^2}\left(\eta_H(f_{200}+h_L f_{110}+h_L^2 f_{020})+\eta_H^2(f_{201}+h_L f_{111}+h_L^2 f_{021})\right).
\]
At order $\eta_H^{1}$, using the symmetry under the exchange of $h_L$ and $h_H$, we have $f_{110}=f_{101},f_{020}=f_{002}$, which can be calculated by expanding $f_{10}$ (\ref{eq:ExactResult})  and $f_{00}$ (\ref{eq:f00}) in terms of $\eta_H$. So the only unknown at this order is $f_{200}$, which means that we only need to solve the $h_{1,2}$ null-state equation at this order to get this term. The result is 
\begin{align}
f_{200}=&\frac{1728(z^2-1)\left(\zeta(3)-\text{Li}_3(1-z)\right)}{z^2}+\frac{288\text{Li}_{2}(z)(7(z-2)z-12(z-1)\log(1-z))}{z^{2}}\nonumber\\
& -\frac{1728(z-2)\text{Li}_{3}(z)}{z}-\frac{144 (z-1) \log ^2(1-z) (6 (z+1) \log (z)-7z+7)}{z^2}+1128\nonumber\\
&+\frac{12 \left(24 \pi ^2 \left(z^2-1\right)+(z-2) z\right) \log (1-z)}{z^2}+\frac{288 (z-2) (z-1)^2 \log ^3(1-z)}{z^3}.\label{eq:f200}
\end{align}
 We have also checked that these results do satisfy the $h_{1,3}$ and $h_{1,4}$ null-state equations.
 
At order $\eta_H^{2}$, only $f_{021}=f_{012}$ can be determined by expanding the result we already have (that is, $f_{01}$), and we need to solve the $h_{1,2}$ and $h_{1,3}$ null-state equations at this order to get $f_{201}$ and $f_{111}$. These results are complicated and given in appendix \ref{appendixfmnk}.

Using the symmetry $f_{ijk}=f_{ikj}$, we can also determine the liner $\eta_H$ terms at order $1/c^3$ by just using the the $h_{1,2}$ null-state equation. At this order,
\[
\text{log}\CV\supset \frac{h_L\eta_H}{c^3}(f_{300}+h_L f_{210}+h_L^2 f_{120}+h_L^3 f_{030}).
\]
Since $f_{210}=f_{201}$, $f_{120}=f_{102}$ and $f_{030}=f_{003}$,  only $f_{300}$ cannot be obtained by expanding the results we already have, that's why we only need the $h_{1,2}$ null-state equation. These results are also given in appendix \ref{appendixfmnk}.


\subsection{Integral Formulas from the Coulomb Gas}
\label{sec:CoulombGasMethod}

As we mentioned in the previous sections, computation of $f_{mn}$ at higher orders becomes extremely technically challenging, because upon the substitution $h_L \to h_{r,s}$ one needs to solve a differential constraint equation of order $rs$. However, an integral representation of the solutions to constraint equations such as \eqref{eq:NullDiffEqh21} are known, thanks to the Coulomb gas formalism \cite{Dotsenko:1984nm, Dotsenko:1984ad, DiFrancesco:1997nk}. This method makes it possible to write down explicit expressions for all $f_{mn}$ in terms of multiple elementary integrals.

Explicitly, the vacuum block component of $\frac{\langle \CO_{1,s}(0) \CO_{1,s}(z)\CO_H(1)\CO_H (\infty)\rangle}{\langle \CO_{1,s}(0) \CO_{1,s}(z)\rangle\langle\CO_H(1)\CO_H (\infty)\rangle}$, where $\CO_{1,s}$ is a light degenerate operator, is given by the following integral representation: 
\begin{align}\label{eq:IR V1s}
\tilde{\Vcal}_{1,s}(z)
=N_{1,s}\left( \prod_{i=1}^{s-1}\int_0^1\! d w_i  \right)
(1-z)^{(s-1)\beta_H}
e^{{\cal I}_{1,s}}\;,
\end{align}
where the action $\CI_{1,s}$ is
\begin{align}
\CI_{1,s}=\sum_{i=1}^{s-1}
\left\{
\frac{s-1}{b^2}\log\big[w_i (1-w_i)\big]-2\beta_H \log(1-z w_i)
\right\}
-\frac{2}{b^2}\sum_{1\le i < j \le s-1}\log(w_i-w_j)\;.
\end{align}
with $\beta_H$ given by \eqref{eq:alphaH}.
We have also introduced a normalization factor $N_{1,s}$ such that ${\tilde \CV}_{1,s}(0) = 1$. 
Notice that $N_{1,s}$ is independent of $h_H$.  Perturbatively in $b$, it is given by
\be\label{eq:N1s}
N_{1,s}(b)=1+\frac{4(s-1)^2-3(s-1)(s-2)}{2b^2}+\CO(b^{-4})\;.
\ee
In the limit $b\to \infty$ with fixed $\beta_H$, we can expand the integrand of \eqref{eq:IR V1s} in $1/b$:
\begin{align}\label{eq: V1s expansion}
{\tilde \CV}_{1,s}(z)&=N_{1,s}(b)
(1-z)^{(s-1)\beta_H}
\int_0^1\!
\left(\prod_{i=1}^{s-1}d w_i
\right) (1-z w_i)^{-2\beta_H}\nonumber\\
&\times
\sum_{k=0}^\infty\frac{1}{k!b^{2k}}
\left(\sum_{i=1}^{s-1}(s-1)K_i-\sum_{1\le i<j\le s-1}2U_{ij}\right)^k\;.
\end{align}
To lighten the notation, we denote
\begin{align}
K_i&=\log(w_i(1-w_i))\;,\quad U_{ij}=\log|w_i-w_j|\;.
\end{align}

In the rest of this section, we will show how to extract various $f_{mn}$ from the integral \eqref{eq: V1s expansion}. The general strategy  is very simple. Recall that we postulated the ansatz of the vacuum block to be
\begin{align}\label{eq: ansatz tV}
\tilde{\cal V}_{h_H,h_L,0,c}(z) & =z^{2h_L}\exp\left[h_L \sum_{n,m=0}^{\infty}\left(\frac{1}{c}\right)^{m}\left(\frac{h_{L}}{c}\right)^{n}f_{mn}\left(\eta_H,z\right)\right].
\end{align}
When we set $h_L=h_{1,s}=\frac{1-s}{2}+\frac{1-s^2}{4b^2}$ in the above ansatz and compare it with \eqref{eq: V1s expansion}, we can read off the $f_{mn}$ functions.


\subsubsection{Leading Order at Large $c$}

Let us begin by computing the well-known $c=\infty$ heavy-light vacuum block as a warm-up. Upon substitution $h_L\to h_{1,s}$, Eq.~\eqref{eq: ansatz tV} in leading order in $b$ is simply $z^{2h_L}\exp(\frac{1-s}{2}f_{00})$. Denoting $X_{1,s}$ the $(s-1)$ dimensional integral in \eqref{eq: V1s expansion}, the comparison implies
\begin{align}
z^{2h_L}e^{\frac{1-s}{2}f_{00}}=N_{1,s}^{(0)} X_{1,s}^{(0)}
=
\prod_{i=1}^{s-1}\left[(1-z)^{\beta_H}
\int_0^1\!d w_i (1-z w_i)^{-2\beta_H}
\right]\;,
\end{align}
where the superscript denotes the powers in $\frac{1}{b^2}$, e.g.~$X_{1,s}=X_{1,s}^{(0)}+\frac{1}{b^2}X_{1,s}^{(1)}+\dots$.
From the above equation one immediately obtains that 
\be\label{eq:f00 2}
f_{00}=-2 \log
\left(\frac{(1-z)^{\beta_H}-(1-z)^{1-\beta_H}}{(1-2\beta_H) }
\right)\;.
\ee
Noting that in large $b$ limit $\beta_H\to \frac{1-\sqrt{1-24\eta_H}}{2}+\Ocal(b^{-2})$, we recognize the above equation in agreement with \eqref{eq:f00}.

\subsubsection{Expansion at Order $1/c$}
Now we arrive at the sub-leading order in $c$. They are two functions, $f_{10}$ and $f_{01}$, to be determined at this order. The comparison of \eqref{eq:LargeCAnsatz} with \eqref{eq: V1s expansion} yields
\begin{align}
\frac{1-s^2}{4}(f_{00}+2\log z)+\frac{1-s}{2}\left(\frac{f_{10}}{6}+\frac{f_{01}}{6}\frac{1-s}{2}\right)&=\frac{N_{1,s}^{(1)} X_{1,s}^{(0)}+N_{1,s}^{(0)} X_{1,s}^{(1)}}{N_{1,s}^{(0)} X_{1,s}^{(0)}}\;.
\end{align}
In the above equation, $N_{1,s}^{(0)}$ and $N_{1,s}^{(1)}$ on the RHS are obtained from \eqref{eq:N1s}, while $X_{1,2}^{(0)}$ and $X_{1,2}^{(1)}$ are represented by elementary integrals. Staring at \eqref{eq: V1s expansion}, one finds that 
\begin{align}
X_{1,s}^{(0)}=\left(\int_w\mathbf{1}\right)^{s-1}\;,\quad 
X_{1,s}^{(1)}=X_{1,s}^{(0)}\left((s-1)^2
\frac{\int_{w_1}K_1}{\int_{w_1}\mathbf{1}}-(s-1)(s-2)\frac{\int_{w_1}\int_{w_2}U_{12} }{[\int_{w_1}\mathbf{1}]^2}
\right)\;.
\end{align}
Here we have used the abbreviation 
\begin{align}
\int_{w_i} f(w_1,\dots, w_n)\equiv(1-z)^{\beta_H}
\int_0^1\! d w_i (1-z w_i)^{-2\beta_H}f(w_1,\dots, w_n)\;.
\end{align}
Combining all these pieces of information, one can easily solve for $f_{10}$ and $f_{01}$: 
\begin{align}\label{eq:f10f01}
f_{10}&=-18-6(f_{00}+2\log z)-12\frac{\int_{w_1}\int_{w_2}U_{12} }{[\int_{w_1}\mathbf{1}]^2}\;,\nn\\
f_{01}&=12+6(f_{00}+2\log z)+24\frac{\int_{w_1}K_1}{\int_{w_1}\mathbf{1}}-24\frac{\int_{w_1}\int_{w_2}U_{12} }{[\int_{w_1}\mathbf{1}]^2}\;,
\end{align}
where $f_{00}$ is given by \eqref{eq:f00 2}. Now what remains to be computed are the two integrals in the expressions above.  After some cumbersome but straightforward algebra, one has
\begin{align}
\frac{\int_{w_1}K_1}{\int_{w_1}\mathbf{1}}&=
\frac{
\int_0^1\! d w (1-z w)^{-2\beta_H}\log\big|w (1-w)\big|}{\int_0^1\! d w (1-z w)^{-2\beta_H}}\label{F1}\\
&=\bigg(
\frac{-\alpha  H_{-\alpha }+\alpha  \log \left(-\frac{1}{z}\right)-2+(1-z)^{\alpha } \left(2+\alpha
    \left(\psi ^{(0)}(\alpha )+\log \left(\frac{z}{z-1}\right)+\gamma
   \right)\right)}{\alpha}\nn\\
&+\frac{\, _2F_1\left(1,\alpha ;\alpha +1;\frac{1}{1-z}\right)z+\frac{\alpha}{1-\alpha}
   (1-z)^{\alpha } \, _2F_1\left(1,1;2-\alpha ;\frac{1}{z}\right)}{\alpha z}
\bigg)\left(1-(1-z)^{\alpha}\right)^{-1}\;,\nn\\
\frac{\int_{w_1}\int_{w_2}U_{12} }{[\int_{w_1}\mathbf{1}]^2}&=
\frac{
\int_0^1\!d w_1
\int_0^1\!d w_2 \big[(1-z w_1)(1-z w_2)\big]^{-2\beta_H}\log|w_1-w_2|
}
{\left(\int_0^1\! d w (1-z w)^{-2\beta_H}\right)^2}\label{F2}\\
&=\bigg[\frac{i \pi  \alpha +8 (1-z)^{\alpha }-2 \alpha  \log (z)+(1-z)^{2 \alpha } \left(i \pi 
   \alpha -2 \alpha  \log(\frac{z}{1-z})-1\right)-1}{2 \alpha}\nn\\
&+\frac{ (1-z)^{2 \alpha } \left(B(1-z,-\alpha ,0)-3 B\left(\frac{1}{1-z},\alpha
   ,0\right)\right)+ B\left(\frac{1}{1-z},-\alpha ,0\right)-3 B(1-z,\alpha ,0)}{2}\nn\\
&+\frac{\pi  \cot (\pi  \alpha )-2 H_{\alpha }+(1-z)^{2 \alpha } \left(\pi  \cot (\pi  \alpha )-2
   H_{\alpha }\right)}{2}\bigg]
\left(1-(1-z)^{\alpha}\right)^{-2}\;,\nn
\end{align}
where $B(x,\beta,0)=\frac{x^{\beta}{}_{2}F_{1}(1,\beta,1+\beta,x)}{\beta}$
is the incomplete Beta function,  $H_{n}$ is the
harmonic function, $\gamma$ is the Euler gamma constant, $\psi(x)=\frac{\Gamma'(x)}{\Gamma(x)}$ is the digamma function and the parameter $\alpha$ is related to the Hawking temperature by $\alpha \equiv \sqrt{1-24\eta_H}=2\pi i T_H$. Having \eqref{F1} and \eqref{F2} plugged into the expression of $f_{10}$ and $f_{01}$ (\ref{eq:f10f01}), it is straightforward to show that they match the results obtained in \cite{Fitzpatrick:2015dlt}, which are also given in appendix \ref{appendixfmnk}.

One can easily continue this procedure to higher orders in the $1/c$ expansion, but for brevity we spare the reader the details, since the lengthy $1/c^2$ and $1/c^3$ results have already been given in section \ref{sec:VacBlockThirdOrderDiff} and in appendix \ref{appendixfmnk}.


\section{All-Orders Resummations in the Lorentzian Regime}
\label{sec:AllOrdersResummation}

Our main focus in this section is to understand how the large $c$ vacuum Virasoro block behaves in the Lorentzian regime.  More specfically, we are interested in the behavior of the block after the argument $z$ is analytically continued across the branch cut emanating from $ z = 1$ and then taken to small values of $|z|$ on the second sheet.   The behavior of CFT correlators in this regime has interesting implications for causality  \cite{Hartman:2015lfa, Hartman:2016dxc, Li:2015itl, Hofman:2016awc} and a fascinating interpretation in terms of chaos \cite{Shenker:2013pqa, Roberts:2014ifa, Shenker:2014cwa, Maldacena:2015waa, JoeRuelle,  Fitzpatrick:2016thx, Perlmutter:2016pkf}.  

\begin{figure}[ht!]
\begin{center}
\includegraphics[width=0.7\textwidth]{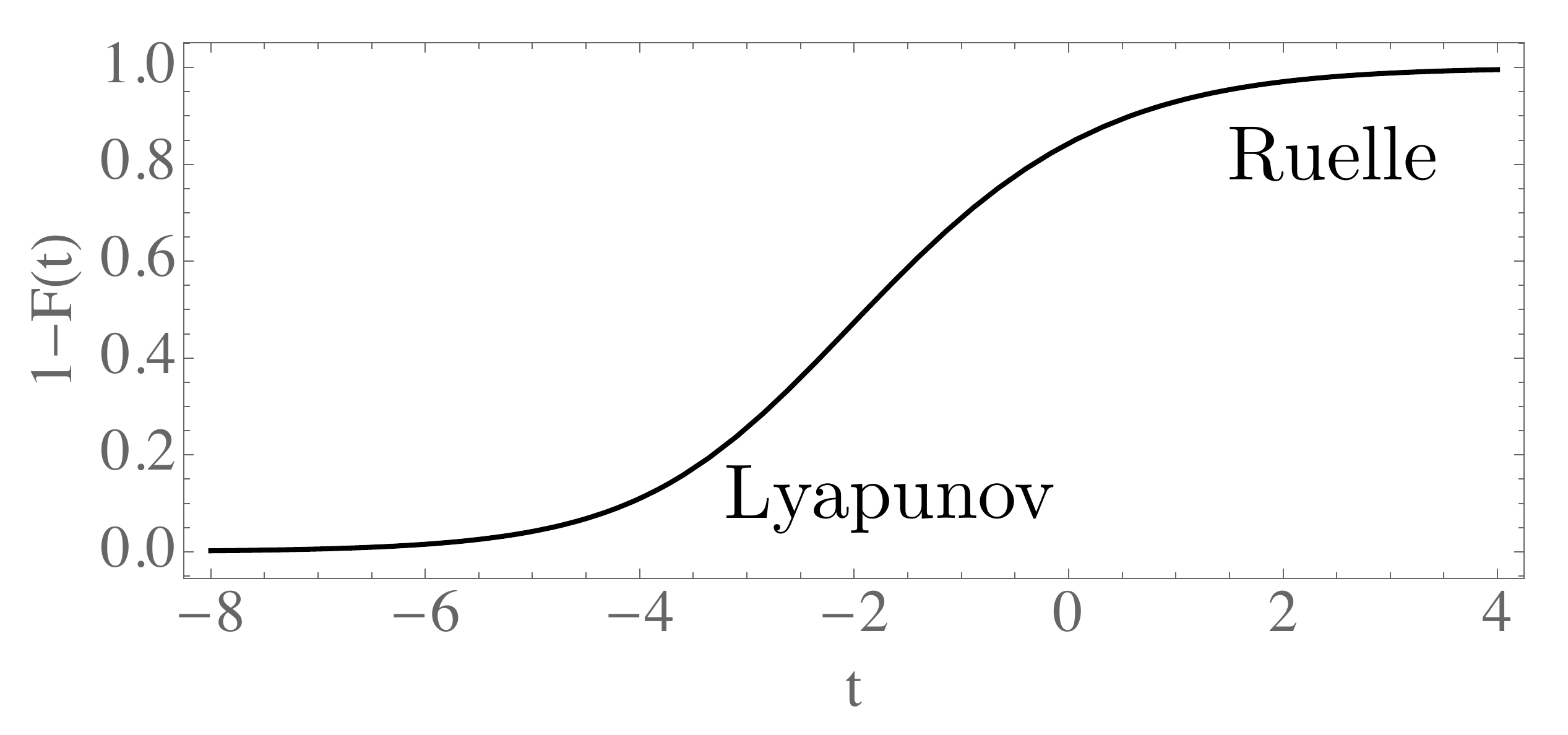}
\caption{Plot of the behavior of  $1-F(t)$ as a function of time $t$ in the limit $c\rightarrow \infty$ with $c z$ fixed, with $h_L = h_H = \frac{1}{2}$.  $F(t)$ is absolute value of the out-of-order correlator $\frac{\< \CO_L \CO_H \CO_L \CO_H\>_\beta}{\<\CO_L \CO_L\> \<\CO_H \CO_H\>}$, and $t \equiv -\log( c z/6)$.  The initial ``Lyapunov'' growth and the later ``Ruelle'' decay are labeled as in \cite{JoeRuelle}.  We have plotted only the contribution of an approximation to the vacuum Virasoro block, but the result has the qualitative features expected of the full correlator. }
\label{fig:LyapRuelle}
\end{center}
\end{figure}

In this section we will show that quantum corrections to the Lyapunov exponent resum to all orders, and that one can also resum the full $\frac{1}{cz}$ expansion in order to obtain an interpolation between the early onset of chaos and late time effects associated with thermalization.  These are the Lyapunov and Ruelle regions of figure \ref{fig:LyapRuelle}.  We refer the reader to \cite{Fitzpatrick:2016thx} for a pertinent review of chaotic correlators and Lyapunov exponent bounds in the context of CFT$_2$ at large central charge.

\subsection{Resummation of $\frac{1}{c} \log z$ Effects}

\label{sec:logresum}

Consider the Virasoro vacuum  block in a large $c$ expansion with external dimensions fixed.  In a $1/c$ expansion, the leading correction near $z\sim 0$ after analytically continuing around the branch cut emanating from $z = 1$ is of the form 
\be
F(z) \approx 1 - \frac{48 i \pi h_L h_H}{c z} + \dots
\label{eq:firstLorentzSingTerm}
\ee 
The first term comes from the vacuum itself, while the second term is due entirely to the exchange of a single quasi-primary stress tensor or `graviton' state along with its global conformal descendants.  The quantity $F$ is the contribution of the vacuum block to the out of time order correlator  $\< \CO_H \CO_L \CO_H  \CO_L \>_\beta$ in a thermal background, normalized by the $\< \CO_H \CO_H \>\< \CO_L \CO_L \>$ ; it is plotted in figure \ref{fig:LyapRuelle}.

As $z$ decreases towards $0$, the $1/c$ correction grows like $z^{-1}$ and becomes increasingly important.  Similarly, higher order terms in $1/c$ can become important at sufficiently small $z$ as well. In this subsection we will show how to resum one set of contributions that grow large at small $z$, namely the terms that are leading logs in the $1/c$ expansion.  That is, we will see that terms of the form $(1/z c)(\log(z)/c)^n$ appear exactly in the combination
\be
\frac{A}{c z^{1+\gamma/c}} =  \frac{A}{z c}\sum_{n=0}^\infty \frac{1}{n!} \left( \frac{-\gamma \log(z)}{c} \right)^n,
\label{eq:expandedexp}
\ee
 with constants $A=-48 i \pi h_L h_H,\gamma=12$.  We provide another derivation of this resummation in appendix \ref{app:DirectLogResummation}. We also checked the coefficients of these terms by analytically continuing the $f_{m00}$ functions given in appendix \ref{appendixfmnk} directly to the second sheet up to and including $1/c^3$ corrections. These effects provide a quantum correction \cite{Fitzpatrick:2016thx} to the Lyapunov exponents that characterize the early onset of chaos.
  
 For degenerate external operators, there is a particularly transparent way of understanding this logarithmic resummation, because only a finite number of Virasoro blocks appear in any channel.  The crucial point is that passing through the branch cut in $z$ simply reshuffles one linear combination of blocks into a different linear combination.\footnote{One way of understanding this is that crossing symmetry $z \rightarrow 1-z$ acts as a linear operator that changes blocks in one channel into blocks in the other channel.  In the other channel, taking $z$ around $1$ acts on each block by simply  introducing a phase $(1-z)^{h_I} \rightarrow e^{2 \pi i h_I} (1-z)^{h_I}$ given by the weight $h_I$ of the corresponding primary operator.  Transforming back to the original channel again acts with the inverse of the first linear operator, producing a linear combination of blocks in the original channel. }   In other words, on the second (Lorentzian) sheet, the vacuum block is equal to a sum of degenerate blocks evaluated on the first (or Euclidean) sheet.

For the degenerate operator $\CO_{1,s}$, the operators in the $\CO_{1,s} \times \CO_{1,s}$ OPE are degenerate operators 
$\CO_{1,p}$ with $p=1, 3, \dots, 2s-1$. These have dimension 
\be
h_{1,p} &=& 
 -\frac{1}{2}(p-1) \left( 1 + \frac{1}{2} b^{-2} (p+1) \right).
\ee
where we recall $c \approx 6b^2 \gg 1$.
So for a given value of $s$, analytic continuation of $z$ around $1$ transforms the vacuum block into a linear combination of terms of the form
\be
{\tilde \CV}_{(1,s)}(z) &\sim& \sum_{q=0}^{s-1} \frac{c_q(h_{1,s},h_H)}{b^{2q}} \frac{1}{z^{q (1+ b^{-2} (q+1))}} f_q(h_{1,s}, h_H, z) ,
\label{eq:genSS}
\ee
where $q\equiv \frac{(p-1)}{2}$  and the $f_q(z) \sim 1 + \CO(z)$ parts of the blocks have a regular series expansion around $z\sim 0$.
In the above, $c_q$ and $f_q$ are functions of $b$ as well but we have factored out explicit powers of $b^{-2}$ so that they have a finite limit at $b\rightarrow \infty$. The reason this prefactor of $b^{-2q}$ must be present is that $c_q$ vanishes up to $\CO(b^{-2q+2})$, by the following argument.  If we expand at large $b$, we know that the $b^{-2q+2}$ term is a $(q-1)$-th order polynomial in $h_{1,s}$, and therefore given by $q$ coefficients.\footnote{These coefficients are functions of $z$ and $h_H$.} These coefficient can be fixed by looking at the OPE of the $q$ degenerate operators $\{ \CO_{1,s}\}_{1\le s \le q}$.  From the above description of the $\CO_{1,s} \times \CO_{1,s}$ OPE, we know that none of the operators $\{ \CO_{1,s} \}_{1 \le s \le q}$ contains the $\CO_{1,2q+1}$ operator, therefore this operator does not appear at $\CO(b^{-2q+2})$ or lower.  But, $c_q$ is just  the OPE coefficient for the $\CO_{1,2q+1}$ operator; therefore the lowest order where it appears is $b^{-2q}$.


Now, to see explicitly the behavior of leading logs, sub-leading logs, sub-sub-leading logs, etc, we can simply expand in large $b$ and look for terms of order $(b^{-2} \log(z))^n$, $b^{-2} (b^{-2} \log(z))^n$, etc.  Logarithms manifestly arise only from expanding an exponent of $z$ in the above expression, so any term of the form 
\be
(b^{-2} )^m (b^{-2} \log(z))^n
\ee
must come from expanding an exponent $n$ times after expanding the prefactor up to $m$-th order.  There are manifestly no terms with $m=0$.  Terms with $m=1$ must clearly come from the first term, $q=1$, and are the leading logs. Consequently, we immediately see that all these leading logs arise from the expansion of the term
\be
\frac{c_1(h_{1,s},h_H) f_1(h_{1,s}, h_H,z)}{b^2 z^{1+2 b^{-2}}}
\ee
and thus manifestly just resum back to this form.  This result holds for all values of $s$.  Since we expect that the vacuum block $\CV$ is analytic in $h_L$, and because this result obtains for $h_L = h_{1,s}$ for all $s$, we expect that it also holds if we analytically continue to general $h_L$.   The correction to the power-law in the denominator is
\be
1\rightarrow 1 + 2 b^{-2} = 1 + \frac{12}{c} + \CO(c^{-2}),
\ee
proving equation (\ref{eq:expandedexp}) with $\gamma=12$.  This also provides a quick alternative check of the magnitude and sign of the correction to this (Lyapunov) exponent.

The above considerations also make it easy to understand the effect of sub-leading logs, sub-sub-leading logs, etc. For instance terms with $m=2$ must come from either the first term or the second term in (\ref{eq:genSS}), and therefore are of the form
\be
&& z^{2h_{1,s}} \<  \CO_H(\infty) \CO_H(1) \CO_{1,s}(z) \CO_{1,s}(0) \> \supset \nn\\
 && \qquad \sum_{n=0}^{\infty} (b^{-2})^2 \left( \frac{\left[ c_1 f_1(z)\right]_{\CO(b^{-2})} (-2 b^{-2} \log(z))^n}{z n!} 
+ \frac{\left[ c_2 f_2(z) \right]_{\CO(b^0)} (-6 b^{-2}\log(z))^n}{z^2 n!} \right) .\nn\\
\ee
It is easy to expand in large $b$ to obtain similar higher order results.

\subsection{Resumming Leading Singularities in $\frac{1}{cz}$}
\label{sec:leadingsing}

Resumming the leading logarithms tells us something about the functional form of the large $c$ expansion, but because of the power-law singularities $\sim (c z)^{-n}$, the leading logs never dominate the behavior of the blocks. In this subsection, we will derive and resum  the leading $(c z)^{-n}$ singularities,  which do give the dominant behavior at small $z$ in the limit $c\rightarrow \infty$ with $c z $ fixed. 

\begin{figure}[ht!]
\begin{center}
\includegraphics[width=0.45\textwidth]{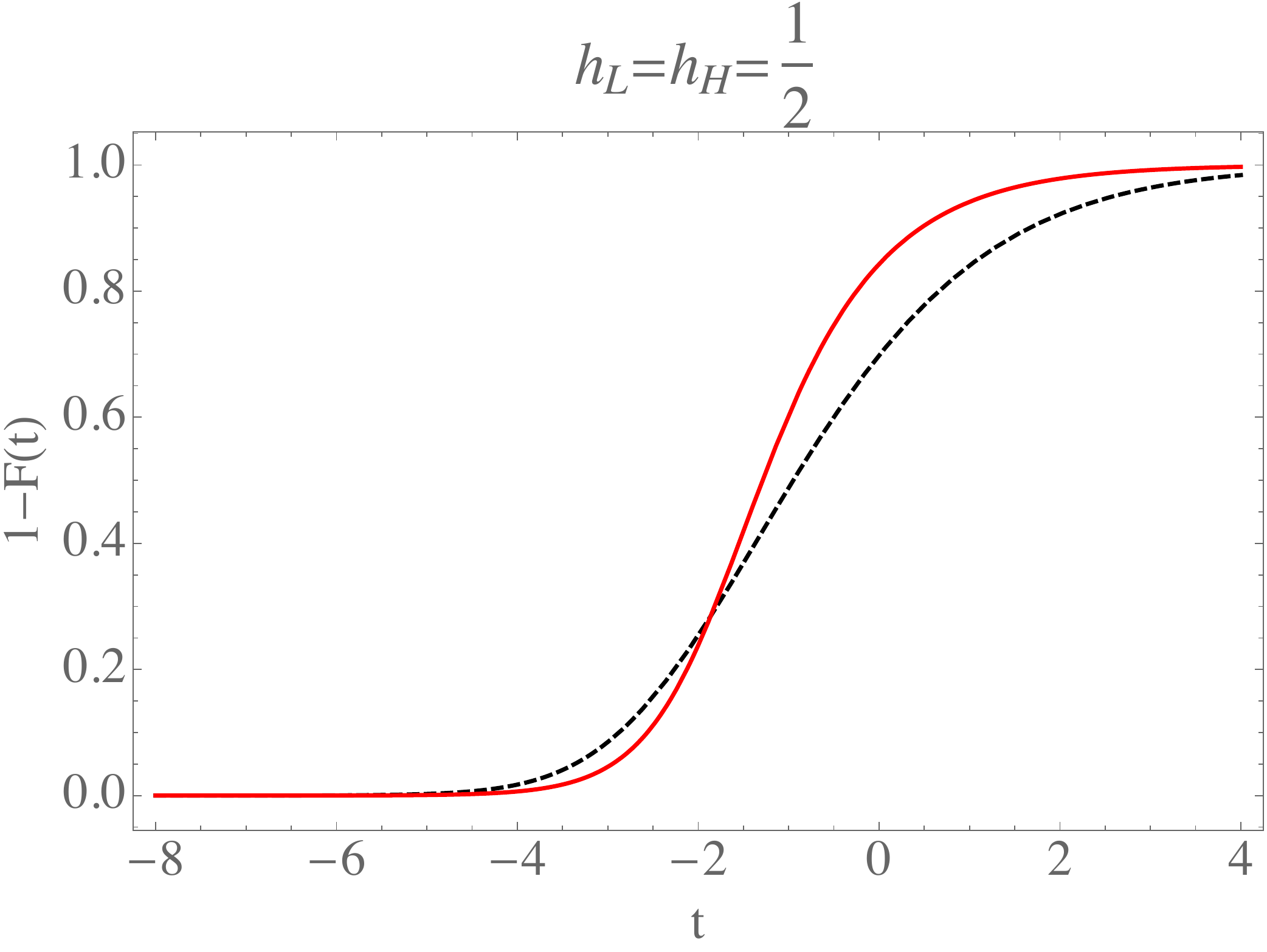}
\includegraphics[width=0.45\textwidth]{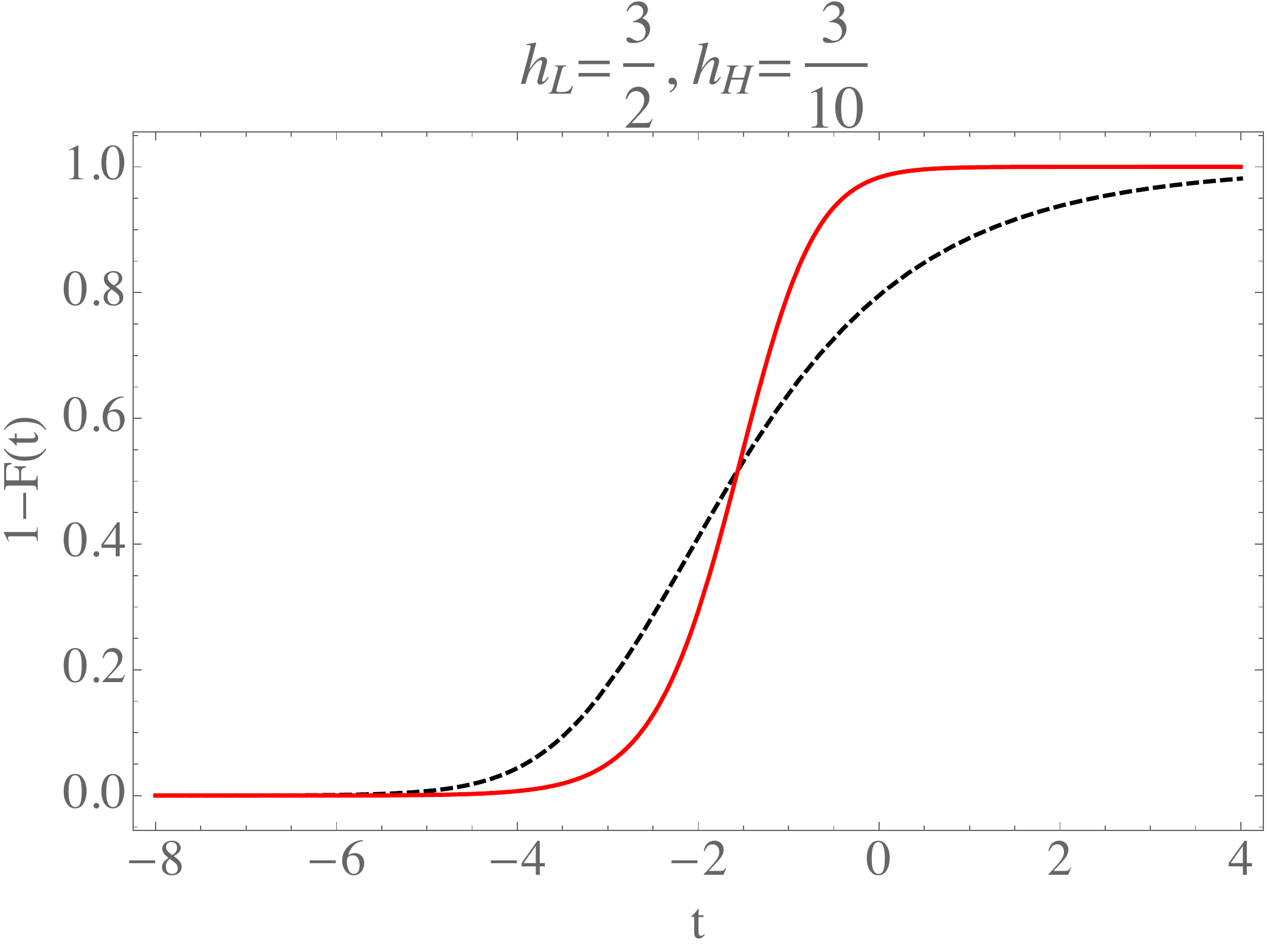}
\caption{Plots comparing the exact behavior from eq. (\ref{eq:finalresumA})  (black, dashed) for  $1-F(t)$  in the limit $c\rightarrow \infty$ with $c z$ fixed, to the heavy-light approximation (\ref{eq:heavylightlorentz}) (red, solid). {\it Left}: $h_L = h_H = \frac{1}{2}$, {\it Right}: $h_L = \frac{3}{2}, h_H = \frac{3}{10}$.   $F(t)$  and $t  $ are  as in figure \ref{fig:LyapRuelle}.  Note that both curves only include contributions from the vacuum block, neglecting double-trace operators which could affect an AdS$_3$ calculation. }
\label{fig:LyapRuelleComparisons}
\end{center}
\end{figure}

The arguments in the previous subsection already provide a significant amount of information on the coefficients of these singularities in equation (\ref{eq:genSS}): they are polynomials in $h_L$ and $h_H$ of order $n$, they have to vanish when $h_L$ is a degenerate operator $h_{1,s}$ with $s \le n$, and they have to be symmetric in $h_L \leftrightarrow h_H$.  This in fact completely determines the coefficients $c_q(h_L, h_H) $ of equation (\ref{eq:genSS}) up to an $h_H, h_L$-independent prefactor:
\be
c_q(h_L, h_H) &=& a_q (2h_L)_q (2h_H)_q,
\label{eq:cqform}
\ee
where $a_q$ depends only on $q$ and not on $h_H$ or $h_L$.  To obtain its value, we just need to calculate it for some chosen $h_H$, in the limit $c \rightarrow \infty$.  A convenient choice is $h_H = \eta_H c$ fixed, followed by $\eta_H$ small, since in that case we know from the form of the heavy-light blocks in the $c\rightarrow \infty$ limit that, on  the second sheet \cite{Roberts:2014ifa, Fitzpatrick:2016thx}, the vacuum block is \cite{Fitzpatrick:2014vua} 
\be
z^{2h_L} \CV(z) &\approx &\left( \frac{1}{1- \frac{24 i \pi h_H}{c z} } \right)^{2h_L} .
\label{eq:heavylightlorentz}
\ee
Series expanding in $1/c$, we can read off the $c_q$ coefficients in this limit and determine the prefactor $a_q$, with the result\footnote{Note that since the approximation (\ref{eq:heavylightlorentz}) retains some of the $h_H$-dependence and all of the $h_L$-dependence of the coefficients $c_q$ in its $1/c$ series expansion, this also provides a non-trivial consistency check of equation (\ref{eq:cqform}).}
\be
c_q(h_L, h_H) &=& \frac{ (2 i \pi)^q (2h_H)_q (2h_L)_q}{q!}
\ee
Substituting these coefficients back into the sum over singular terms
\be \label{eq:1overcz}
\sum_{q=0}^\infty \frac{c_q(h_L,h_H)}{b^{2q} z^q},
\ee
we see that the sum on $q$ is an asymptotic series, ie it has zero radius of convergence.  One can nevertheless  Borel resum it:
\be
B(t) &=& \sum_{q=0}^\infty \frac{c_q t^q}{q!} = {}_2 F_1(2h_L, 2h_H, 1, 2 i \pi t).
\ee
Performing the Borel integral $\int_0^\infty e^{-t} B(\frac{t}{b^2 z} )dt$, we obtain a relatively compact expression for the resummation of the leading singular terms:
\be
\lim_{c \rightarrow \infty \atop c z \textrm{ fixed}} (z^{2h_L}) \CV(z) &=&
  G \left(h_H, h_L, \frac{i c z}{12\pi} \right) + G \left( h_L,h_H, \frac{i c z}{12\pi} \right)
  \label{eq:finalresumA}
\ee
where
\be  
G(h_1,h_2,x) &\equiv & (x )^{2h_1} (2h_2)_{-2h_1} ~ {}_1F_1(2h_1,1+2h_1-2h_2, x).
\label{eq:finalresum}
\ee
This might be compared with the integral formulas from \cite{Shenker:2014cwa} derived from AdS physics.
As one might  expect, we see that the singular terms all resum into something that shuts down at $z \sim 0$. The two terms above decay like $z^{2h_H}$ and  $z^{2h_L}$, respectively.  Suggestively, these exponents would naively correspond to the contributions from a $\CO_H \CO_H$ double-trace operator and  a $\CO_L \CO_L$ double-trace operator.  This is closely related to the fact that if one takes the expression for the vacuum block in the heavy-light limit
\be
\CV \propto \left( \frac{\pi T_H}{\sin^2 (\pi T_H (t+i \phi)}\right)^{2h_L}\left( \frac{\pi \bar{T}_H}{\sin^2 (\pi \bar{T}_H (t-i \phi)}\right)^{2h_L}
\ee
and promotes it to a periodic function of $\phi$ (which the full correlator must be) by adding all its images under $\phi \rightarrow \phi + 2 \pi n$, then this generates additional contributions in the conformal block decomposition that behave like double-trace operators in the $\CO_L \CO_L$ OPE.  It is interesting that, unlike the global conformal blocks, the Virasoro conformal blocks thereby ``know'' about double-trace operator contributions in the same channel $\CO_L \CO_L \rightarrow \CO_H \CO_H$ as the vacuum.

Adopting the nomenclature of \cite{JoeRuelle}, the above expression interpolates between the  ``Lyapunov'' regime, where $c$ is large with $c z$ fixed and large, and the  ``Ruelle'' regime, where $c$ is large with $c z$ fixed and small. 
For $h_H=h_L$, the expression simplifies somewhat:
\be
 \lim_{h_H\rightarrow h_L} G(h_L,h_H,x)+ G(h_H,h_L,x)  
 &=& x^{2 h_L} U(2 h_L,1,x) .
\ee
where $U(a,b,x)$ is a confluent hypergeometric function.\footnote{For $b \notin \mathbb{Z}$, 
\be
U(a,b,x) &=& \frac{\Gamma(b-1)}{\Gamma(a)} z^{1-b} {}_1F_1(a-b+1, 2-b, x) + \frac{\Gamma(1-b)}{\Gamma(a-b+1)} {}_1F_1(a,b,x)
\ee}
It is particularly simple at $h_L=1/2$, since $U(1,1,x) = e^x \Gamma(0,x)$.  
In figure \ref{fig:LyapRuelle}, we have plotted the resulting behavior for the correlator (only including the vacuum block contributions) interpolating between the Lyapunov and Ruelle regime for $h_L=h_H=\frac{1}{2}$.  In figure \ref{fig:LyapRuelleComparisons}, we compare the  behavior  of the vacuum block with that of the approximate formula (\ref{eq:heavylightlorentz}) from the heavy-light limit.  Although all of these plots only include vacuum block contributions, they seem to agree with qualitative expectations for the behavior of the full correlator.

We make one final comment on the relation of this result to the heavy-light limit.  One open question has been whether or not taking the heavy-light limit, then analytically continuing around $z\sim 1$, and finally taking $c$ large with $h_L, h_H,$ and $c z$ fixed is the same as simply analytically continuing the exact Virasoro block and then taking the limit $c$ large with $h_L, h_H,$ and $c z$ fixed.  So far, all indications are that these different orders of limits do commute for the $\CO(1/c)$ singular term (\ref{eq:firstLorentzSingTerm}), which was the main interest of \cite{Roberts:2014ifa}, but in the above we see explicitly that they do not commute for most other terms.  In particular,  taking the heavy-light limit followed by small $h_H$ completely discards the contribution in (\ref{eq:finalresum}) that decays like $z^{2h_H}$, since by inspection we see that (\ref{eq:heavylightlorentz})
 contains only the $\sim (c z)^{2h_L}$ piece at small $c z$. This is perhaps not so surprising, since the full result has to be symmetric under $h_L \leftrightarrow h_H$, but taking the heavy-light limit breaks this symmetry and makes the $\CO(z^{2h_H})$ contributions become formally non-perturbative $\sim e^{2 \eta_H c \log(z)}$. By contrast, by working out the exact coefficient of the leading singularities, we have kept the $h_H \leftrightarrow h_L$ symmetry at all stages of the computation.

\section{Heavy-Light Super-Virasoro Vacuum Blocks at Large $c$}
\label{sec:SuperVirasoro}

Similar to the case of non-supersymmetric CFTs that we have being discussing so far, in two-dimensional superconformal theories (SCFTs) there are degenerate operators whose correlators satisfy super null-state differential equations. In this section, we will use these super null-state equations to calculate the large $c$ heavy-light  super-Virasoro vacuum block for these degenerate operators, and then analytically continue the result to obtain the super-Virasoro vacuum block for operators with general conformal dimensions. Specifically, we will focus on the holomorphic part of the  Neveu-Schwarz (NS) sector of 2d $\mathcal{N}=1$ \cite{FRIEDAN198537,QIU1986205,Bershadsky:1985dq,Eichenherr:1985cx,Fuchs:1986ew,CHAOSHANGHUANG199381}  and $\mathcal{N}=2$  \cite{DiVecchia:1985ief,DiVecchia:1986cdz,Boucher:1986bh, Kiritsis:1987np,Mussardo:1988av,Dorrzapf:1994es} SCFTs (see e.g. \cite{Blumenhagen:2013fgp} for a review of these theories). Previously, the $\mathcal{N}=1$ super-Virasoro blocks in NS sector have been studied using recursion relations \cite{Belavin:2006zr,Hadasz:2006qb,Hadasz:2007nt}, while those of $\mathcal{N}=2$ are less investigated \cite{Belavin:2012qh,Lin:2015wcg}.  

\subsection{The $\mathcal{N}=1$ Super-Virasoro Vacuum Block}

\subsubsection{Brief review of 2d $\mathcal{N}=1$ SCFTs}
In the $\CN=1$ super-space, a point is denoted by $Z\equiv\left(z,\theta\right)$, where $\theta$ is a Grassmann variable. A primary superfield $\Phi_h (Z)$ of conformal dimension $h$  can be expanded in terms of $\theta$ as $\Phi_h (Z)=\phi_h(z)+\theta\psi_{h+\frac{1}{2}}(z)$, where $\phi_h(z)$ and $\psi_{h+\frac{1}{2}}(z)$ are two component fields with conformal dimension $h$ and $h+\frac{1}{2}$, respectively . In the NS sector, the  energy-momentum superfield $\CT(Z)$, which has conformal dimension $3/2$, can be expanded around the origin as 
\be
\mathcal{T}(Z)=\sum_{r\in \mathbb{Z}+\frac{1}{2}} \frac{1}{2z^{r+3/2}}G_r+\theta\sum_{n\in \mathbb{Z}} \frac{1}{z^{n+2}}L_n,
\ee
where the fermionic generators $G_r$ are the supersymmetry generators and the bosonic generators $L_n$ are Virasoro generators. The (anti-)commutation relations between these generators are:
\begin{equation}\label{eq:n=1commutation}
\begin{aligned}
\left[L_{n},L_{m}\right]=&\left(n-m\right)L_{n+m}+\frac{c}{12}\left(n^{3}-n\right)\delta_{n+m,0},\\
\left\{ G_{r},G_{s}\right\} =&2L_{r+s}+\frac{c}{3}\left(r^{2}-\frac{1}{4}\right)\delta_{r+s,0}, \\
\left[L_{n},G_{r}\right]=&\left(\frac{n}{2}-r\right)G_{n+r}, \qquad m,n\in \mathbb{Z};\ r,s\in\mathbb{Z}+\frac{1}{2}.
\end{aligned}
\end{equation}
The singular terms in the OPE of $\mathcal{T}(Z_1)$ and $\Phi(Z_2)$ are
\[
\mathcal{T}(Z_1)\Phi(Z_2)\sim \frac{h\theta_{12}}{Z_{12}^2}\Phi(Z_2)+\frac{1}{2Z_{12}}D_2\Phi(Z_2)+\frac{\theta_{12}}{Z_{12}}\partial_2\Phi(Z_2),
\]
where $Z_{ij}=z_{ij}-\theta_{i}\theta_{j}, z_{ij}=z_{i}-z_{j} \text{ and } D_{i}=\partial_{\theta_{i}}+\theta_{i}\partial_{z_{i}}.$ Descendant superfields are obtained by acting on a primary with $L_{-n}$ and $G_{-r}$ for $n,r>0$. From the above OPE, one can derive that correlation functions with one descendant superfield can be written in terms of a differential operator acting on the correlation functions with only primary superfields \cite{CHAOSHANGHUANG199381} via
\begin{equation}
\left\langle (L_{-n}\Phi_1)(Z_1)X\right\rangle=\CL_{-n}\left\langle (\Phi_1)(Z_1)X\right\rangle,\quad \left\langle (G_{-r}\Phi_1)(Z_1)X\right\rangle=\mathcal{G}_{-r}\left\langle (\Phi_1)(Z_1)X\right\rangle,
\end{equation}
where $X=\Phi_2(Z_2)\cdots\Phi_N(Z_N)$ is an assembly of primary superfields, and $\Phi_i$ has conformal dimension $h_i$. These two super-differential operators are 
\begin{align}
\CL_{-n}&=-\sum_{i=2}^{N}Z_{i1}^{-n}[(1-n)(h_i+\frac{1}{2}\theta_{i1}D_i)+Z_{i1}\partial_{z_i}]\left\langle\Phi_1(Z_1)X\right\rangle\nonumber,\\
\mathcal{G}_{-r}
 &=-\sum_{i=2}^{N}Z_{i1}^{-(r+\frac{1}{2})}\left[(2r-1)h_i\theta_{i1}+Z_{i1}(D_i-2\theta_{i1}\partial_{z_i})\right]\left\langle\Phi_1(Z_1)X\right\rangle. \label{eq:descendantdifferential}
\end{align}


$N$-point functions of the superfields $F_N\equiv\left\langle\Phi_1(Z_1)\Phi_2(Z_2)\cdots\Phi_N(Z_N)\right\rangle$ should be invariant under the global superconformal transformations generated by
$L_{\pm1},L_{0},G_{\pm\frac{1}{2}}$, which leads to the superconformal Ward identities \cite{Fuchs:1986ew}:
\begin{equation}
\begin{aligned}
&L_{-1} : \sum_{i=1}^N \partial_{z_i}F_N=0,\quad  G_{-\frac{1}{2}},G_{\frac{1}{2}}: \sum_{i=1}^N (\partial_{\theta_i}-\theta_{i}\partial_{z_{i}})F_N=\sum_{i=1}^N(2h_i\theta_{i}+z_i(\theta_i\partial_{z_i}-\partial_{\theta_i}))F_N=0,\\
 &L_0 : \sum_{i=1}^N (2z_i\partial_{z_i}+2h_i+\theta_i\partial_{\theta_i})F_N=0,\quad  L_{1}: \sum_{i=1}^N (z_i^2\partial_{z_i}+z_i(2h_i+\theta_i\partial_{\theta_i}))F_N=0.
\end{aligned}
\end{equation}
Due to these constraints, the two-point function is fixed (up to normalization) to be 
\begin{equation}\label{eq:n=1twopt}
\left\langle \Phi_1\left(Z_{1}\right)\Phi_2\left(Z_{2}\right)\right\rangle =\frac{1}{Z_{21}^{2h_1}}\delta_{h_1,h_2}=\left(\frac{1}{z_{21}^{2h_1}}+\theta_1\theta_2\frac{-2h_1}{z_{21}^{2h_1+1}}\right)\delta_{h_1,h_2}.
\end{equation}
where each term on the RHS corresponds to a two-point function of the component fields.

\subsubsection{$\mathcal{N}=1$ Super-Virasoro Vacuum Blocks at Leading Oder}
The heavy-light super-Virasoro vacuum block $\CV_{\Phi_L\Phi_L\Phi_H\Phi_H}$ is the contribution to the heavy-light four-point function  $\left\langle \Phi_{L}\left(Z_{1}\right)\Phi_{L}\left\langle Z_{2}\right\rangle \Phi_{H}\left\langle Z_{3}\right\rangle \Phi_{H}\left(Z_{4}\right)\right\rangle$ from an irreducible representation of the superconformal algebra whose highest weight state is the vacuum $\left|0\right\rangle$. In the following calculation, we will take the heavy-light limit, meaning that 
$$\eta_H\equiv\frac{h_H}{c}, h_L \ \text{ fixed  as} \ \ c\rightarrow \infty.$$
Our result of this part is  $\CV_{\Phi_L\Phi_L\Phi_H\Phi_H}$ given in (\ref{eq:n=1ansatz}) with $f_{h_L}$ and $g_{h_L}$ given in (\ref{eq:n=1fhL}) and (\ref{eq:n=1ghL}).

As the four-point function, the super-Virasoro vacuum block  $\CV_{\Phi_L\Phi_L\Phi_H\Phi_H}$ also satisfies the superconformal Ward identities. There are eight coordinate variables (four Grassmann even and four Grassmann odd) in  $\CV_{\Phi_L\Phi_L\Phi_H\Phi_H}$ and it satisfies five global superconformal Ward identities, which means that there are only three independent superconformal invariants, two Grassmann even and one Grassmann odd.  The two Grassmann even invariants that we choose are \cite{QIU1986205}
\begin{equation}
x_{0} \equiv \frac{Z_{12}Z_{34}}{Z_{13}Z_{24}},\ \ x_{1} \equiv \frac{Z_{14}Z_{23}}{Z_{13}Z_{24}}-\left(1-x_{0}\right).
\end{equation}
It is easy to verify that $x_{1}^{2}=0 $ and superformal Ward identities fix  $\CV_{\Phi_L\Phi_L\Phi_H\Phi_H}$ (which is Grassmann even) to be of the following general form:
\begin{equation}
\CV_{\Phi_L\Phi_L\Phi_H\Phi_H} =\frac{1}{Z_{21}^{2h_{L}}Z_{34}^{2h_{H}}}\left[f_{h_L}\left(x_{0}\right)+x_{1}g_{h_L}\left(x_{0}\right)\right] .\label{eq:n=1ansatz}
\end{equation}

The conformal dimensions of the degenerate fields in the NS sector of an $\mathcal{N}=1$
SCFTs can be parameterized by 
\begin{equation}
h_{r,s}=\frac{\left[\left(m+2\right)r-ms\right]^{2}-4}{8m\left(m+2\right)}, \  c = \frac{3}{2}-\frac{12}{m\left(m+2\right)}\qquad r,s\in \mathbb{Z}^+;\  r-s\in 2\mathbb{Z}.
\end{equation}
and the corresponding null-state is at level $\frac{rs}{2}$.
The first non-trivial null state ($r=1,s=3$) is 
\begin{equation}
\left(\frac{2}{2h_{1,3}+1}L_{-1}G_{-1/2}-G_{-3/2}\right)\left|\Phi_{1,3}\right\rangle =0,
\end{equation}
with $h_{1,3}=-\frac{1}{2}-\frac{3}{c}+\CO\left(1/c^2\right)$ in the large $c$ limit.
If $\Phi_{L}=\Phi_{1,3}$ in the heavy-light four-point function  $\left\langle \Phi_{L}\Phi_{L} \Phi_{H} \Phi_{H}\right\rangle$, then 
\begin{equation}
\left\langle \left(\frac{2}{2h_{13}+1}L_{-1}G_{-1/2}-G_{-3/2}\right)\Phi_{1,3}\left(Z_{1}\right)\Phi_{1,3}\left(Z_{2}\right)\Phi_{H}\left(Z_{3}\right)\Phi_{H}\left(Z_{4}\right)\right\rangle =0.
\end{equation}
Using (\ref{eq:descendantdifferential}), we get a null-state equation satisfied by the four-point function, which also satisfied by the super-Virasoro vacuum block. Simplifying this equation using the superconformal Ward identities ($\CL_{-1}$ becomes just $\partial_{z_1}$ and $\mathcal{G}_{-1/2}$ becomes just $D_1=\partial_{\theta_{1}}+\theta_{1}\partial_{z_{1}}$), we find
\[
\left\{ \frac{2\partial_{z_{1}}\left(\partial_{\theta_{1}}+\theta_{1}\partial_{z_{1}}\right)}{2h_{1,3}+1}+\sum_{i=2}^4\left[Z_{i1}^{-1}\left(\partial_{\theta_{i}}-\theta_{i}\partial_{z_{i}}+2\theta_{1}\partial_{z_{i}}\right)+2h_{i}\theta_{i1}Z_{i1}^{-2}\right]\right\} \CV_{\Phi_{1,3}\Phi_{1,3}\Phi_H\Phi_H}=0
.\]
This is a super-differential equation with two unknown function $f(x_1)$ and $g(x_1)$. To solve it, we can expand it in terms of $\theta_i$s and require that all the coefficients of $\theta_i$s equal to zero. First, we can send the $z_i$s to  $\left(0,z,1,\infty\right)$,
in which case, $x_{0}$ and $x_{1}$ become 
\begin{align*}
x_{0}&\rightarrow  z+\theta_{1}\theta_{2}-z\theta_{1}\theta_{3},\\
x_{1}&\rightarrow  \theta_{1}\theta_{2}-\theta_{1}\theta_{3}+\theta_{2}\theta_{3}.
\end{align*}
Expanding the super-differential equation in terms of $\theta_i$s, we get two differential equations from the coefficients of $\theta_1$ and $\theta_2$ (differential equations from coefficients of other $\theta_i$s are dependent with these two). In the large $c$ limit, with $h_{1,3}=-\frac{1}{2}-\frac{3}{c}+\CO\left(1/c^2\right)$ and $\eta_H=\frac{h_H}{c}$  fixed, the leading order ($c^0$) of these two equations are\footnote{
	For later reference, the exact differential equations are 
	\begin{align*}
	f''_{h_{1,3}}+\frac{2 (3-z) h_{1,3}+3 z-1}{2 (z-1) z}f'_{h_{1,3}}-\frac{\left(2 h_{1,3}+1\right)  h_H}{(z-1)^2}f_{h_{1,3}}+\frac{2 h_{1,3}+1}{2 (z-1) z}g_{h_{1,3}}&=0\\
    f''_{h_{1,3}}+\frac{(6-4 z) h_{1,3}+2 z-1}{2 (z-1) z}f'_{h_{1,3}}+g'_{h_{1,3}}+\frac{z-2 (z-2) h_{1,3}}{2 (z-1) z}g_{h_{1,3}}&=0
	\end{align*}} 
\begin{align*}
(z-1)^{2}\left(zf_{h_{1,3}}''(z)+2f_{h_{1,3}}'(z)\right)+z\eta_H f_{h_{1,3}}(z)&=  0,\\
z\left(f_{h_{1,3}}''(z)+g_{h_{1,3}}'(z)\right)+2f_{h_{1,3}}'(z)+g_{h_{1,3}}(z)&=  0.
\end{align*}
Solving these equations and fixing the constants of integration to match the expansion of the vacuum block in terms of small $z$, we find
\begin{align}
f_{h_{1,3}}\left(z\right) & =z^{-1}e^{-\frac{1}{2}f_{00}\left(z\right)},\\
g_{h_{1,3}}\left(z\right) & =\frac{1}{z}-\frac{f_{h_{1,3}}\left(z\right)}{z}-f_{h_{1,3}}'\left(z\right). \label{eq:n=1gh13}
\end{align}
where $f_{00}\left(z\right)$ is defined in equation (\ref{eq:f00}). These solutions only apply to $h_{L}=h_{1,3}$ in the large $c$ limit. But the appearance of $f_{00}(z)$ in $f_{h_{1,3}}(z)$ gives us some hints for how to analytically continue to find $f_{h_L}(z)$ for general $h_L$, which is what we are going to do in the following. After getting $f_{h_L}(z)$, we can use it to obtain $g_{h_L}(z)$ without using the null-state equations.

Expanding both sides of the vacuum block of the superfields (\ref{eq:n=1ansatz}) in terms 
of $\theta_{i}$s and matching the coefficients of $\theta_i$s, we can obtain relations between vacuum blocks of the component fields\footnote {These vacuum blocks are normalized such that the first term of the small $z$ expansion of a vacuum block $\mathcal{V}_{\CO_L(0)\CO_L(z)\CO_H(1)\CO_H(\infty)}$ is $\left\langle\CO_L(0)\CO_L(z)\right\rangle$.} and the functions $f_{h_L}(z)$ and $g_{h_L}(z)$ :
\begin{align}
\CV_{\phi_L\phi_L\phi_H\phi_H}=&z^{-2h_L}f_{h_L}\left(z\right)\label{eq:n=1fourphi},\\
\CV_{\psi_L\psi_L\phi_H\phi_H}=&-z^{-2h_{L}}\left(f_{h_L}'(z)-\frac{2h_{L}f_{h_L}(z)}{z}+g_{h_L}(z)\right)\label{eq:n=1psipsiphiphi}.
\end{align}
where $\CV_{\phi_L\phi_L\phi_H\phi_H}$ is from the term without $\theta_i$ in it and $\CV_{\psi_L\psi_L\phi_H\phi_H}$ is from the coefficient of $\theta_1\theta_2$. In (\ref{eq:n=1psipsiphiphi}), the minus sign in front is due to the fact that $\theta$ anti-commutes with $\psi$. Using equation (\ref{eq:n=1fourphi}) for $h_{1,3}=-\frac{1}{2}+\CO\left(1/c\right)$ in the leading large $c$ limit, we have $\CV_{\phi_{1,3}\phi_{1,3}\phi_H\phi_H}= zf_{h_{1,3}}=e^{-\frac{1}{2}f_{00}(z)}$,  which suggests that for general $h_L$, we should have
\begin{equation}
  \mathcal{V}_{\phi_{L}\phi_{L}\phi_{H}\phi_{H}}=e^{h_{L}f_{00}\left(z\right)}.\label{eq:n=1fourphiblock}
\end{equation}
Using equation (\ref{eq:n=1fourphi}) again, we have
\begin{equation} \label{eq:n=1fhL}
f_{h_{L}}\left(z\right)=z^{2h_L}e^{h_{L}f_{00}\left(z\right)}.
\end{equation}
From equation (\ref{eq:n=1fourphiblock}), one can see that the super-Virasoro vacuum block $\CV_{\phi_L \phi_L \phi_H \phi_H }$ in $\mathcal{N}=1$ SCFTs is the same as the vacuum block in non-susy CFTs at leading order of the large $c$ limit. We explain in detail why this is true in appendix \ref{appendixn=1}, but the basic point is that in this limit, only the pure Virasoro generators contribute to the sum over intermediate states in this block.  

To get $g_{h_L}(z)$, we need to know $\mathcal{V}_{\psi_{L}\psi_{L}\phi_{H}\phi_{H}}$. At leading order of the large $c$ limit, the only difference between $\mathcal{V}_{\psi_{L}\psi_{L}\phi_{H}\phi_{H}}$ and $\mathcal{V}_{\phi_{L}\phi_{L}\phi_{H}\phi_{H}}$ (up to normalization)  is that the conformal dimensions of the light operators are different ($h_{\phi_L}=h_{L},h_{\psi_{L}}=h_{L}+\frac{1}{2}$). \footnote{This point can be seen from the commutation relations of the Virasoro generators with these component fields \eqref{eq:n=1commutator}, and at leading order of large $c$ limit, only Virasoro generators contribute to these two vacuum blocks.}
Since we know $\mathcal{V}_{\phi_{L}\phi_{L}\phi_{H}\phi_{H}}=e^{h_{L}f_{00}\left(z\right)}$, we can immediately see that
\begin{equation}
\mathcal{V}_{\psi_{L}\psi_{L}\phi_{H}\phi_{H}}=2h_Le^{\left(h_{L}+\frac{1}{2}\right)f_{00}\left(z\right)},
\end{equation}
where the prefactor $2h_L$ is due to our convention of the vacuum block and can be read off from the two-point function of superfields (\ref{eq:n=1twopt}). Equating the above vacuum block to (\ref{eq:n=1psipsiphiphi}), we find
\begin{equation}\label{eq:n=1ghL}
g_{h_{L}}\left(z\right)=-2h_{L}z^{2h_{L}}e^{\left(h_{L}+\frac{1}{2}\right)f_{00}\left(z\right)}+\frac{2h_{L}f_{h_{L}}\left(z\right)}{z}-f_{h_{L}}'\left(z\right).
\end{equation}
One can check that setting $h_L=-\frac{1}{2}$ gives us back $g_{h_{1,3}}$ (\ref{eq:n=1gh13}). 

Having the expressions for $f_{h_L}$ and $g_{h_L}$, we can restore their argument to $x_0$, then other super-Virasoro vacuum blocks of the component fields can be read off from the expansion of $\CV_{\Phi_L\Phi_L\Phi_H\Phi_H}$ (\ref{eq:n=1ansatz}) in terms of the $\theta_i$ variables.

\subsection{The $\mathcal{N}=2$ Super-Virasoro Vacuum Block}
\subsubsection{Brief Review of 2d $\mathcal{N}=2$ SCFTs}
In the $\mathcal{N}=2$ superspace, a point is denoted by $Z\equiv\left(z,\theta,\overline{\theta}\right)$, where $z$ is the usual complex coordinate, while $\theta$ and $\overline{\theta}$ are two Grassmann coordinates. The  energy-momentum superfield can be expanded as 
\begin{equation}
\mathcal{J}\left(Z\right)=J\left(z\right)+\theta\overline{G}\left(z\right)-\overline{\theta}G\left(z\right)+\theta\overline{\theta}2T\left(z\right).
\end{equation}
where $J(z)$ is the $U(1)$ $R$-current. The mode expansions are defined in the usual way
\[
J(z)=\sum_{n\in\mathbb{Z}}\frac{J_n}{z^{n+1}},\ G(z)=\sum_{r\in \mathbb{Z}+\frac{1}{2}}\frac{G_r}{z^{r+\frac{3}{2}}},\ \overline{G}=\sum_{r\in\mathbb{Z}+\frac{1}{2}}\frac{\overline{G}_r}{z^{r+\frac{3}{2}}},\ T(z)=\sum_{n\in\mathbb{Z}}\frac{L_n}{z^{n+2}}.
\]
The full $\mathcal{N}=2$  superconformal algebra of these generators takes the following form: 
\begin{equation}
\begin{aligned}
\left[L_{m},L_{n}\right]=&\left(m-n\right)L_{m+n}+\frac{c}{12}\left(m^{3}-m\right)\delta_{m+n,0},\\
\left[L_{m},G_{r}\right]=&\left(\frac{m}{2}-r\right)G_{m+r},\quad \left[L_{m},\overline{G}_{r}\right]=\left(\frac{m}{2}-r\right)\overline{G}_{m+r},\\
\left[J_{m},J_{n}\right]=&\frac{c}{3}m\delta_{m+n,0},\quad \left[L_{m},J_{n}\right]=-nJ_{m+n},\label{eq:n=2commutation}\\
\left[J_{m},G_{r}\right]=&G_{m+r},\quad \left[J_{m}, \overline{G}_{r}\right]=-\overline{G}_{m+r},\\
\left\{ G_{r},\overline{G}_{s}\right\} =&2L_{r+s}+\left(r-s\right)J_{r+s}+\frac{c}{3}\left(r^{2}-\frac{1}{4}\right)\delta_{r+s,0}\\
\left\{ G_{r},G_{s}\right\} =&\left\{ \overline{G}_{r}, \overline{G}_{s}\right\} =0, \qquad m,n\in \mathbb{Z};\  r,s \in \mathbb{Z}+\frac{1}{2}.
\end{aligned}
\end{equation}
A superfield $\Phi\left(Z\right)$ can be expanded in terms of $\theta$ and $\overline{\theta}$ as 
\begin{equation}
\Phi_{h}^{q}\left(Z\right)=\phi_{h}^{q}\left(z\right)+\theta\overline{\psi}_{h+\frac{1}{2}}^{q-1}\left(z\right)+\overline{\theta}\psi_{h+\frac{1}{2}}^{q+1}\left(z\right)+\theta\overline{\theta}\lambda_{h+1}^{q}\left(z\right),\label{eq:n=2superfield}
\end{equation}
where the superscripts and subscripts 
are the conformal dimensions and $U\left(1\right)$ charges of the component fields. The OPE of $\CJ(Z_1)$ and $\Phi(Z_2)$ is 
\[
\CJ(Z_1)\Phi(Z_2)\sim \frac{2h\theta_{12}\overline{\theta}_{12}}{Z_{12}^2}\Phi(Z_2)+\frac{\theta_{12}D_2-\overline{\theta}_{12}\overline{D}_2}{Z_{12}}
\Phi(Z_2)+\frac{2\theta_{12}\overline{\theta}_{12}}{Z_{12}}\partial_{z_2}\Phi(Z_2)+\frac{q}{Z_{12}}\Phi(Z_2).
\]
where the super derivatives and super-translationally invariant distance are 
\begin{equation}
D_{i}=\partial_{\theta_i}+\overline{\theta}_i\partial_{z_i},\ \ \overline{D}_{i}=\partial_{\overline{\theta}_i}+\theta_i\partial_{z_i}, \ \ Z_{ij}\equiv z_{ij}-\theta_{i}\overline{\theta}_{j}-\overline{\theta}_{i}\theta_{j}, \label{eq:n=2superderivative}
\end{equation}
with $z_{ij}=z_{i}-z_{j}$, $\theta_{ij}=\theta_{i}-\theta_{j}$
and $\overline{\theta}_{ij}=\overline{\theta}_{i}-\overline{\theta}_{j}$.

The highest weight states in the NS sector are characterized by their
eigenvalues under $L_{0}$ and $J_{0}$ :
\begin{equation}
L_{0}\left|\Phi\right\rangle =h\left|\Phi\right\rangle ,J_{0}\left|\Phi\right\rangle =q\left|\Phi\right\rangle,
\end{equation}
and they satisfy
\begin{equation}
L_{n}\left|\Phi\right\rangle =J_{n}\left|\Phi\right\rangle =G_{r}\left|\Phi\right\rangle =\overline{G}_{r}\left|\Phi\right\rangle =0,\quad \text{for } n,r>0.
\end{equation}
Acting on primary superfield with $L_{-n},G_{-r},\overline{G}_{-r},J_{-n} (n,r>0)$, we get the descendent superfields.
Using the OPE of $\CJ$ and $\Phi$, one can show that the correlation function with one descendant superfield can be written in terms of a super-differential operator acting on a correlation function with only primary fields:
\begin{equation}\label{eq:n=2differentialoperator}
\begin{aligned}
\left\langle (L_{-n}\Phi_1)(Z_1)X\right\rangle&=\CL_{-n}\left\langle\Phi_1(Z_1)X\right\rangle, &\left\langle (J_{-n}\Phi_1)(Z_1)X\right\rangle=\CJ_{-n}\left\langle\Phi_1(Z_1)X\right\rangle,\\
\left\langle (G_{-r}\Phi_1)(Z_1)X\right\rangle&=\mathcal{G}_{-r}\left\langle\Phi_1(Z_1)X\right\rangle,  &
\left\langle (\overline{G}_{-r}\Phi_1)(Z_1)X\right\rangle=\overline{\mathcal{G}}_{-r}\left\langle\Phi_1(Z_1)X\right\rangle.
\end{aligned}
\end{equation}
where $X=\Phi_2(Z_2)\cdots\Phi_N(Z_N)$ is an assembly of primary fields with conformal dimension $h_i$ and $U(1)$ charge $q_i$. These super-differential operators are \cite{Dorrzapf:1994es}
\begin{equation}\label{eq:n=2DiffOperators}
\begin{aligned}
\mathcal{L}_{-n}&=-\sum_{i=2}^N Z_{i1}^{-n}  [(1-n)(h_i+\frac{1}{2}\theta_{i1}D_{i}+\frac{1}{2}\overline{\theta}_{i1}\overline{D}_{i})+Z_{i1}\partial_{z_{i}}-\frac{q_i}{2}\theta_{i1}\overline{\theta}_{i1}Z_{i1}^{-1}n(1-n)],\\
\mathcal{J}_{-n} &=-\sum_{i=2}^N Z_{i1}^{-n}\left(\overline{\theta}_{i1}\overline{D}_{i}-\theta_{i1}D_{i}+2\theta_{i1}\overline{\theta}_{i1}\partial_{z_{i}}+q_i-2h_i\theta_{i1}\overline{\theta}_{i1}nZ_{i1}^{-1}\right),\\
\mathcal{G}_{-r}  &=-\sum_{i=2}^N Z_{i1}^{-r-\frac{1}{2}}[(r-\frac{1}{2})\theta_{i1}(2h_i+q_i+\overline{\theta}_{i1}\overline{D}_{i})+Z_{i1}(\overline{D}_{i}-2\theta_{i1}\partial_{z_{i}})],\\
\overline{\mathcal{G}}_{-r}  &=-\sum_{i=2}^N Z_{i1}^{-r-\frac{1}{2}}[(r-\frac{1}{2})\overline{\theta}_{i1}(2h_i-q_i+\theta_{i1}D_{i})+Z_{i1}(D_{i}-2\overline{\theta}_{i1}\partial_{z_{i}})].
\end{aligned}
\end{equation}

N-point correlation functions of the primary superfields should be invariant under the global super-conformal transformations generated by $L_{\pm1},L_0, J_0, G_{\pm\frac{1}{2}},\overline{G}_{\pm\frac{1}{2}}$, which leads to the superconformal Ward identities (\ref{eq:n=2Ward}). These Ward identities completely fix the two-point functions to be 
\begin{align}
\left\langle \Phi_{1}(Z_1)\Phi_{2}(Z_2)\right\rangle &=  \frac{1}{Z_{21}^{2h_1}}e^{q_{2}\frac{\theta_{12}\overline{\theta}_{12}}{Z_{12}}}\delta_{q_{1}+q_{2},0}\delta_{h_{1},h_{2}}\nonumber \\
&=  \left(\frac{1}{z_{21}^{2h_{1}}}+\frac{-q_{2}}{z_{21}^{2h_{1}+1}}\theta_{1}\overline{\theta}_{1}+\frac{-2h_{1}+q_{2}}{z_{21}^{2h+1}}\theta_{1}\overline{\theta}_{2}+\frac{-2h_{1}-q_{2}}{z_{21}^{2h_{1}+1}}\overline{\theta}_{1}\theta_{2}+\frac{-q_{2}}{z_{21}^{2h_{1}+1}}\theta_{2}\overline{\theta}_{2}\right.\nonumber \\
&\quad  \left.+\frac{2h_{1}\left(2h_{1}+1\right)}{z_{21}^{2h_{1}+2}}\theta_{1}\overline{\theta}_{1}\theta_{2}\overline{\theta}_{2}\right)\delta_{q_{1}+q_{2},0}\delta_{h_{1},h_{2}}\label{eq:n=2twopt},
\end{align}
up to a normalization constant. Each term in the above equation corresponds to a two-point function of the component fields. Notice that only the two-point function of the lowest component field $\phi$ is normalized as usual.

\subsubsection{Super Null-State Equations}
The heavy-light super-Virasoro vacuum block $\CV_{\Phi_{L}^{-q_L}\Phi_{L}^{q_L}\Phi_{H}^{-q_H}\Phi_{H}^{q_H}}$ is the contribution to the heavy-light four-point function $\langle{\Phi_{L}^{-q_L}(Z_1)\Phi_{L}^{q_L}(Z_2)\Phi_{H}^{-q_H}(Z_3)\Phi_{H}^{q_H}(Z_4)}\rangle$ from an irreducible representation of the superconformal algebra whose highest weight state is the vacuum $\left|0\right\rangle$. In this paper, we will take the following heavy-light limit:
$$ h_L, q_L, \eta_H\equiv\frac{h_H}{c},\eta_{q}\equiv\frac{q_H}{c}\ \text{fixed  as} \ c\rightarrow \infty.$$ 
Our main result of this part is $\CV_{\Phi_{L}^{-q_L}\Phi_{L}^{q_L}\Phi_{H}^{-q_H}\Phi_{H}^{q_H}}$ given in (\ref{eq:n=2ansatz}), with $F\left(x_{0},x_{1},x_{2},x_{3},x_{4}\right)$ given in (\ref{eq:n=2F}) and the  $g_{i,h_L}$ functions given in next subsection \ref{sec:n=2GeneralhL}.

Superconformal Ward identities fix the vacuum block (and the four-point function) to take the following form \cite{Kiritsis:1987np}
\begin{align}
\CV_{\Phi_{L}^{-q_L}\Phi_{L}^{q_L}\Phi_{H}^{-q_H}\Phi_{H}^{q_H}} & =\frac{1}{Z_{21}^{2h_{L}}Z_{34}^{2h_{H}}}\exp \left(q_L\frac{\theta_{12}\overline{\theta}_{12}}{Z_{12}}+q_H\frac{\theta_{34}\overline{\theta}_{34}}{Z_{34}}\right)F\left(x_{0},x_{1},x_{2},x_{3},x_{4}\right),\label{eq:n=2ansatz}
\end{align}
where $F(x_{0},x_{1},x_{2},x_{3},x_{4})$ is a function of five superconformal invariants
\begin{equation*}
x_{0}=\frac{Z_{12}Z_{34}}{Z_{13}Z_{24}} ,\quad x_{1}=\frac{Z_{14}Z_{23}}{Z_{13}Z_{24}}+x_{0}-1,
\end{equation*}
\begin{equation}
\begin{aligned}
x_{2}&=\frac{\theta_{23}\overline{\theta}_{23}}{Z_{23}}+\frac{\theta_{34}\overline{\theta}_{34}}{Z_{34}}-\frac{\theta_{24}\overline{\theta}_{24}}{Z_{24}},\\
x_{3}&=\frac{\theta_{12}\overline{\theta}_{12}}{Z_{12}}+\frac{\theta_{24}\overline{\theta}_{24}}{Z_{24}}-\frac{\theta_{14}\overline{\theta}_{14}}{Z_{14}},\\
x_{4}&=\frac{\theta_{13}\overline{\theta}_{13}}{Z_{13}}+\frac{\theta_{34}\overline{\theta}_{34}}{Z_{34}}-\frac{\theta_{14}\overline{\theta}_{14}}{Z_{14}}.
\end{aligned}
\end{equation}
It is easy to verify that these super-conformal invariants satisfy the  relations
\[
x_{1}^{3}=0,\qquad x_{2}^{2}=0,\qquad x_{3}^{2}=0,\qquad x_{4}^{2}=0,\qquad  x_{1}x_{2}=x_{1}x_{3}=x_{1}x_{4}=0,
\]
\[
x_{2}x_{3}x_{0}=x_{2}x_{4},\qquad x_{2}x_{3}=x_{2}x_{4}+x_{3}x_{4},\qquad x_{1}^{2}=2x_{2}x_{3}x_{0}\left(1-x_{0}\right).
\]
which means that the most general form of $F\left(x_{0},x_{1},x_{2},x_{3},x_{4}\right)$ can be written as
\begin{equation}\label{eq:n=2F}
F=g_{0,h_L}(x_0)+x_{1}g_{1,h_L}(x_0)+x_{2}g_{2,h_L}(x_0)+x_{3}g_{3,h_L}(x_0)+x_{4}g_{4,h_L}(x_0)+x_{2}x_{3}g_{5,h_L}(x_0).
\end{equation}

The conformal dimensions of the degenerate fields in the NS sector of $\mathcal{N}=2$ SCFTs can be parameterized by\footnote{Besides $h_{r,s}$, there are other degenerate fields whose conformal dimensions can be parameterized by $h_{k} = kq+\frac{1}{2}t(k^2-\frac{1}{4}),k\in\mathbb{Z}+\frac{1}{2}$ and having a null field at level $|k|$, but these will not be used in this paper.} \cite{Dorrzapf:1994es}
\begin{equation}
h_{r,s} =  \frac{r^2-1}{8}t-\frac{rs}{4}+\frac{s^2-1}{8t}-\frac{4q^2-1}{8t},\quad  c =  3-3t\qquad r\in\mathbb{Z}^+; s\in 2\mathbb{Z}^+.
\end{equation}
For each degenerate field with dimension $h_{r,s}$, there is a null-field at level $\frac{rs}{2}$. The first non-trivial null-state ($r=1,s=2$) is :
\begin{equation}
\left[\left(q-1\right)L_{-1}-\left(2h_{1,2}+1\right)J_{-1}+G_{-\frac{1}{2}}\overline{G}_{-\frac{1}{2}}\right]\left|\Phi_{1,2}^q\right\rangle=0,\label{eq:nullstate}
\end{equation}
with $h_{1,2}=\frac{c-3q^{2}}{6-2c}=-\frac{1}{2}+\frac{3\left(q^{2}-1\right)}{2c}+\CO\left(1/c^2\right).$ Notice that the $U(1)$ charge $q$ is a free parameter here. If $h_L=h_{1,2}$ in the heavy-light four-point function, then 
\[
\left\langle\left((-q_L-1)L_{-1}-(2h_{1,2}+1)J_{-1}+G_{-\frac{1}{2}}\overline{G}_{-\frac{1}{2}}\right) \Phi_{1,2}^{-q_L}(Z_1)\Phi_{1,2}^{q_L}(Z_2)\Phi_{H}^{-q_H}(Z_3)\Phi_{H}^{q_H}(Z_4)\right\rangle=0.
\]
Using equations (\ref{eq:n=2differentialoperator}), we get a super-differential equation satisfied by the four-point function, which is also satisfied by the vacuum block  $\CV_{\Phi_{L}^{-q_L}\Phi_{L}^{q_L}\Phi_{H}^{-q_H}\Phi_{H}^{q_H}}$. Simplifying this super-differential equation using the superconformal Ward identities \eqref{eq:n=2Ward} ($\mathcal{L}_{-1}\rightarrow\partial_{z_1}$, $\mathcal{G}_{-\frac{1}{2}}\rightarrow\overline{D}_1$ and $\overline{\mathcal{G}}_{-\frac{1}{2}}\rightarrow D_1$), we find
\begin{equation}
\left[\left(-q_{L}-1\right)\partial_{z_{1}}-\left(2h_{L}+1\right)\mathcal{J}_{-1}+\overline{D}_1D_1\right]\CV_{\Phi_{1,2}^{-q_L}\Phi_{1,2}^{q_L}\Phi_{H}^{-q_H}\Phi_{H}^{q_H}}=0,\label{eq:n=2nullstateequation}
\end{equation}
with $\CJ_{-1}$ given in (\ref{eq:n=2DiffOperators}) and $\overline{D}_1,D_1$ given in (\ref{eq:n=2superderivative}).

To solve this super-differential equation, we can expand it in terms of $\theta_i$s and $\overline{\theta}_i$s  to get six independent differential equations to solve for the six unknown functions\footnote{Again, we send the coordinates $z_{i}$  to $\left(0,z,1,\infty\right)$, in which case, $x_0\rightarrow z+\cdots$, where $\cdots$ represents terms proportional to $\theta_i$, $\overline{\theta}_i$ or their products.} $g_{0,h_{1,2}}(z), \cdots$, $g_{5,h_{1,2}}(z)$. These solutions $g_{i,h_{1,2}}(z)$ only apply to those  vacuum blocks whose light operators are degenerate operators with $h_{L}=h_{1,2}$. To get $g_{i,h_L}(z)$ for general $h_{L}$ we need to analytically continue these solutions, as what we did for the non-susy Virasoro blocks. But in the non-susy case, there was only one unknown function and we already knew its anzatz  for general $h_{L}$ (\ref{eq:LargeCAnsatz}), so things were easier there. Here, we have six $g_{i,h_{1,2}}(z)$ functions and some of them are complicated and hard to know how to analytically continue them. But it turns out that once we solve the differential equation for $g_{0,h_{1,2}}(z)$, then analytically continue the solution to get  $g_{0,h_L}(z)$, we can derive the other $g_{i,h_L}(z)$ functions from it, which will be shown in next subsection \ref{sec:n=2GeneralhL}. The equation  that only involves $g_{0,h_{1,2}}(z)$ is
\begin{equation*}
g_{0,h_{1,2}}''(z)+\left(\frac{6 q_L\eta _q}{z-1}+\frac{2}{z}\right)g_{0,h_{1,2}}'(z)+\frac{6 z \eta _H+3 \eta _q \left(3 z \left(q_L^2-1\right) \eta _q+(z-2) q_L\right)}{(z-1)^2 z}g_{0,h_{1,2}}(z)=0. 
\end{equation*}
The solution is 
\begin{equation}
g_{0,h_{1,2}}(z)=z^{-1}e^{-\frac{1}{2}\tilde{f}\left(z\right)}\left(1-z\right)^{-3\eta_q q_{L}}, \label{eq:n=2g0h12}
\end{equation}
where 
\begin{equation}\label{eq:n=2ftilde}
\tilde{f}\left(z\right)=-(1-\tilde{\alpha})\log\left(1-z\right)-2\log\left(\frac{1-(1-z)^{\tilde{\alpha}}}{\tilde{\alpha}}\right),
\end{equation}
with $\tilde{a}=\sqrt{1-24\eta_H+36\eta_{q}^{2}}$. In the above solution, the constants of integration have been fixed such that the first term in the expansion of $g_{0,h_{1,2}}(z)$ in small $z$ is $1$, which corresponds to the vacuum block.



\subsubsection{Solutions for General $h_{L}$}\label{sec:n=2GeneralhL}
In this subsection, we are going to analytically continue $g_{0,h_{1,2}}$ to get $g_{0,h_L}$, then use it to derive the other $g_{i,h_L}$ functions. Expanding the ansatz (\ref{eq:n=2ansatz}) in terms of $\theta_i$s and $\overline{\theta}_i$s, we can express the vacuum blocks of the component fields in terms of $g_{i,h_L}(z)$ \footnote {These vacuum blocks are normalized such that the first term of the small $z$ expansion of a vacuum block $\mathcal{V}_{\CO_L(0)\CO_L(z)\CO_H(1)\CO_H(\infty)}$ is $\left\langle\CO_L(0)\CO_L(z)\right\rangle$.}:
\begin{align}
\CV_ {\phi_L^{-q_L}\phi_{L}^{q_L}\phi_H^{-q_H}\phi_H^{q_H}} &= z^{-2h_L}g_{0,h_L},\label{eq:n=2fourphiblock}\\
\CV_{\overline{\psi}_L^{-q_L-1}\psi_L^{q_L+1}\phi_H^{-q_H}\phi_H^{q_H}} &= -z^{-(2h_L+1)}\left[\left(q_L-2h_{L}\right)g_{0,h_L}+zg_{1,h_L}+g_{3,h_L}+zg_{0,h_L}'\right],\label{eq:n=2barpsipsiphiphi}\\
\CV_{\psi_L^{-q_L+1}\overline{\psi}_L^{q_L-1}\phi_H^{-q_H}\phi_H^{q_H}} &= z^{-(2h_L+1)}\left[\left(q_L+2h_{L}\right)g_{0,h_L}-zg_{1,h_L}+g_{3,h_L}-zg_{0,h_L}'\right],\label{eq:n=2psibarpsiphiphi}\\
\mathcal{V}_{\phi_L^{-q_L}\lambda_L^{q_L}\phi_H^{-q_H}\phi_H^{q_H}} &=z^{-(2h_L+1)}\left( -q_Lg_{0,h_L}-g_{3,h_L}+\frac{z}{z-1}g_{2,h_L}\right),\label{eq:n=2philambdaphiphi}\\
\mathcal{V}_{\lambda_L^{-q_L}\phi_L^{q_L}\phi_H^{-q_H}\phi_H^{q_H}} &= -z^{-(2h_L+1)}\left(q_Lg_{0,h_L}+g_{3,h_L}+zg_{4,h_L}\right),\label{eq:n=2lambdaphiphiphi}\\
\CV_{\lambda_L^{-q_L}\lambda_L^{q_L}\phi_H^{-q_H}\phi_H^{q_H}} &= z^{-2h_L}\left[g_{0,h_L}''-\frac{4 h_L}{z}\left(g_{0,h_L}'+g_{1,h_L}\right)+\frac{2 h_L \left(2 h_L+1\right)}{z^2}g_{0,h_L}\right.\nonumber\\&\qquad\qquad \left.+2 g_{1,h_L}'+\frac{ 1}{(1-z) z}\left(q_Lg_{2,h_L}+g_{5,h_L}\right)+\frac{q_L}{z}g_{4,h_L}\right].
\label{eq:n=2lambdalambdaphiphi}
\end{align}
The basic idea of these derivations is to derive the vacuum blocks on the LHS,  then solve the above equations to get the functions $g_{i,h_L}(z)$ on the RHS. 

First, the most important function is $g_{0,h_L}(z)$, which is associated with $\CV_ {\phi_L^{-q_L}\phi_{L}^{q_L}\phi_H^{-q_H}\phi_H^{q_H}}$. From equation (\ref{eq:n=2fourphiblock}), for $h_L=h_{1,2}=-\frac{1}{2}+\CO(1/c)$, we have $\CV_{\phi_{1,2}^{-q_L}\phi_{1,2}^{q_L}\phi_H^{-q_H}\phi_H^{q_H}}=zg_{0,h_{1,2}}=e^{-\frac{1}{2}\tilde{f}(z)}(1-z)^{-3\eta_q q_L}$, which suggests that for general $h_L$, we should have 
\begin{equation}
\CV_ {\phi_L^{-q_L}\phi_{L}^{q_L}\phi_H^{-q_H}\phi_H^{q_H}}=e^{h_L\tilde{f}(z)}(1-z)^{-3\eta_q q_L}.
\end{equation}
Indeed, this matches the Virasoro vacuum block for CFT$_2$s with a global  $U\left(1\right)$  symmetry, which have been computed in \cite{Fitzpatrick:2015zha}.
The fact that this super-Virasoro vacuum block  only gets contributions from the Virasoro generators and $U(1)$ generators at leading order of the large $c$ limit can be seen from the commutation relations of these generators with the component field $\phi$, as we explain in the appendix \ref{Appendixn=2Leading}. Using equation (\ref{eq:n=2fourphiblock}) again, we have
\be
g_{0,h_{L}}\left(z\right)=z^{2h_L}e^{h_{L}\tilde{f}\left(z\right)}\left(1-z\right)^{-3\eta_q q_{L}}.\label{eq:n=2g0hL}
\ee

Next, to get $g_{1,h_L}(z)$ and $g_{3,h_L}(z)$, we need to know the blocks $\CV_{\overline{\psi}_L^{-q_L-1}\psi_L^{q_L+1}\phi_H^{-q_H}\phi_H^{q_H}} $ and $\CV_{\psi_L^{-q_L+1}\overline{\psi}_L^{q_L-1}\phi_H^{-q_H}\phi_H^{q_H}}$. At leading order of the large $c$ limit, the only differences between these two blocks and  $\CV_ {\phi_L^{-q_L}\phi_{L}^{q_L}\phi_H^{-q_H}\phi_H^{q_H}}$ are that the conformal dimensions and $U(1)$ charges of the light fields are different (note that the conformal dimensions of $\psi_L$ and $\overline{\psi}_L$ are $h_L+\frac{1}{2}$, while that for $\phi_L$ is $h_L$), which means that we can change
the parameters accordingly in the expression of $\CV_ {\phi_L^{-q_L}\phi_{L}^{q_L}\phi_H^{-q_H}\phi_H^{q_H}}$ to get these two blocks: 
\begin{align*}
\CV_{\overline{\psi}_L^{-q_L-1}\psi_L^{q_L+1}\phi_H^{-q_H}\phi_H^{q_H}} & =  (2h_L-q_L) e^{\left(h_{L}+\frac{1}{2}\right)\tilde{f}\left(z\right)}\left(1-z\right)^{-3\eta_q\left(q_{L}+1\right)}\equiv z^{-2h_L-1}g_{q_L+1}(z),\\
\CV_{\psi_L^{-q_L+1}\overline{\psi}_L^{q_L-1}\phi_H^{-q_H}\phi_H^{q_H}} &= (2h_L+q_L) e^{\left(h_{L}+\frac{1}{2}\right)\tilde{f}\left(z\right)}\left(1-z\right)^{-3\eta_q\left(q_{L}-1\right)}\equiv z^{-2h_L-1} g_{q_L-1}(z).
\end{align*}
where the prefactor $2h_L\mp q_L$ is due to our convention of the definition of the vacuum blocks and can be read off from the two-point function (\ref{eq:n=2twopt}). Equating these two blocks to equations (\ref{eq:n=2barpsipsiphiphi}) and (\ref{eq:n=2psibarpsiphiphi}) respectively, we can solve for  $g_{1,h_{L}}(z)$ and $g_{3,h_{L}}(z)$
\begin{align}
g_{1,h_{L}}=&\frac{1}{2 z}\left(4h_Lg_{0,h_L} -2zg_{0,h_L}'-g_{q_L+1}- g_{q_L-1}\right),\label{eq:n=2g1hL}\\
g_{3,h_{L}}=&\frac{1}{2} \left(-2 q_Lg_{0,h_L}  -g_{q_L+1}+g_{q_L-1}\right).\label{eq:n=2g3hL}
\end{align}

The remaining functions $g_{2,h_L}(z)$, $g_{4,h_L}(z)$ and  $g_{5,h_{L}}(z)$ are related to the vacuum blocks $\mathcal{V}_{\phi_L^{-q_L}\lambda_L^{q_L}\phi_H^{-q_H}\phi_H^{q_H}}$, $\mathcal{V}_{\lambda_L^{-q_L}\phi_L^{q_L}\phi_H^{-q_H}\phi_H^{q_H}}$  and $\mathcal{V}_{\lambda_L^{-q_L}\lambda_L^{q_L}\phi_H^{-q_H}\phi_H^{q_H}}$ (\ref{eq:n=2philambdaphiphi}-\ref{eq:n=2lambdalambdaphiphi}), respectively.
As is shown in appendix \ref{AppendixLambdaDecom}, $\lambda_L^{q_L}$ can be written as descendant fields plus a Virasoro and $U(1)$ primary 
\begin{equation} \label{eq:n=2lambdadecompose}
\lambda_L^{q_L}(z)=\frac{12h^{2}-3q^{2}}{2ch-3q^{2}}(J_{-1}\phi_L^{q_L})(z)+\frac{q(c-6h)}{2ch-3q^{2}}(L_{-1}\phi_L^{q_L})(z)+\tilde{\lambda}_L^{q_L}(z)
\end{equation}
The Virasoro and $U(1)$ primary part $\tilde{\lambda}_L^{q_L}$ has conformal dimension $h_L+1$ and $U(1)$ charge $q_L$, which are the same as $\lambda_L^{q_L}$. Using this decomposition, we can calculate these three vacuum blocks from  $\mathcal{V}_{\phi_L^{-q_L}\phi_L^{q_L}\phi_H^{-q_H}\phi_H^{q_H}}$. Some details for performing these calculations are given in appendix \ref{AppendixLambdaDecom}. Equating these three vacuum blocks to equations (\ref{eq:n=2philambdaphiphi}), (\ref{eq:n=2lambdaphiphiphi}) and  (\ref{eq:n=2lambdalambdaphiphi}), we can solve for $g_{2,h_L}(z)$, $g_{4,h_L}(z)$ and $g_{5,h_L}(z)$. At leading order of the large $c$ limit, these functions are
\begin{align}
g_{2,h_{L}}= &\frac{3 z\eta_q \left(q_L^2-4 h_L^2\right)g_{0,h_L}+2 (z-1) h_Lg_{3,h_L}+(z-1) z q_L g_{0,h_L}'}{2h_L z},\label{eq:n=2g2hL}\\
g_{4,h_{L}}= & g_{2,h_L}-g_{3,h_L},\label{eq:n=2g4hL}\\
g_{5,h_L}=&\frac{(z-1) z}{4}\left(\frac{(z-1) q_L^2}{h_L^2}+4\right)g_{0,h_L}''+ (z-1) \left(2zg_{1,h_L}'-4h_L g_{1,h_L}+q_Lg_{4,h_L}\right)\nonumber\\ 
&-q_Lg_{2,h_L} -\frac{(z-1) \left(6z\eta_q q_L(4h_L^2 -q_L^2)+q_L^2 \left(2 (z-2) h_L-z\right)+16 h_L^3\right)}{4 h_L^2}g_{0,h_L}'\nonumber\\
&+\frac{\left(4 h_L^2-q_L^2\right)^2}{2h_Lz} \left(\frac{ (z^2e^{\tilde{f}}-1) (1-z) \left(2 h_L+1\right)+3 (z-2) z \eta_q q_L}{4h_L^2- q_L^2}+\frac{9\eta_q^2z^2}{2h_L} \right)g_{0,h_L}.\label{eq:n=2g5hL}
\end{align}
with $\tilde{f}$ in $g_{5,h_L}$ given in (\ref{eq:n=2ftilde}). 
One can easily check that setting $h_L=h_{1,2}\simeq-\frac{1}{2}$, these $g_{i,h_L}(z)$ functions will become $g_{i,h_{1,2}}(z)$ and they are indeed the solutions to the null-state equation (\ref{eq:n=2nullstateequation}) at leading order of the large $c$ limit. Restoring the argument of $g_{i,h_L}$ to $x_0$, other super-Virasoro vacuum blocks of the component fields can be read off from the expansion of $\CV_{\Phi_{L}^{-q_L}\Phi_{L}^{q_L}\Phi_{H}^{-q_H}\Phi_{H}^{q_H}}$ in terms of $\theta_i$s and $\overline{\theta}_i$s. We've checked  the first few terms of the expansion of these other blocks in terms of small $z$, and they match the results from the direct calculation of these blocks. \footnote{By direct calculation, we mean to calculate the vacuum block by inserting the vacuum state and its descendants into the four-point function, and then summing over all these contributions.}

\section*{Acknowledgments}
 
We would like to thank David Gross, Tom Hartman, Gary Horowitz, Ami Katz, Don Marolf, Dan Roberts, Edgar Shaghoulian, Douglas Stanford, and Matt Walters for valuable discussions.  ALF is supported by the US Department of Energy Office of Science under Award Number DE-SC-0010025.  JK and HC are supported in part by NSF grants PHY-1316665 and PHY-1454083, and JK is supported by a Sloan Foundation fellowship.  We would also like to thank the GGI in Florence for hospitality as this work was completed.

\appendix
\section{Summary of Corrections to the Vacuum Block} \label{appendixfmnk}
In this appendix, we will list results concerning the large $c$ expansion of the vacuum block. 

At order $c^0$ and $1/c$, $f_{00}$ (\ref{eq:f00}) and $f_{10},f_{01}$ are known in closed form.  The functions $f_{10}$ and $f_{01}$ are known from the work of \cite{Fitzpatrick:2015dlt} to be
\begin{equation}
\begin{aligned}f_{10} & =\frac{\text{csch}^{2}\left(\frac{\alpha t}{2}\right)}{2}\Big[3\left(e^{-\alpha t}B\left(e^{-t},-\alpha,0\right)+e^{\alpha t}B\left(e^{-t},\alpha,0\right)+e^{\alpha t}B\left(e^{t},-\alpha,0\right)+e^{-\alpha t}B\left(e^{t},\alpha,0\right)\right)\\
 & +\frac{1}{\alpha^{2}}+\cosh(\alpha t)\left(-\frac{1}{\alpha^{2}}+6H_{-\alpha}+6H_{\alpha}+6i\pi-5\right)+12\log\left(2\sinh\left(\frac{t}{2}\right)\right)+5\Big]\\
 & -t\frac{\left(13\alpha^{2}-1\right)\coth\left(\frac{\alpha t}{2}\right)}{2\alpha}+12\log\left(\frac{2\sinh\left(\frac{\alpha t}{2}\right)}{\alpha}\right),\\
f_{01} & =6\Big(\text{csch}^{2}\left(\frac{\alpha t}{2}\right)\left[\frac{B(e^{-t},-\alpha,0)+B(e^{t},-\alpha,0)+B(e^{-t},\alpha,0)+B(e^{t},\alpha,0)}{2}\right.\\
 & \left.+H_{-\alpha}+H_{\alpha}+2\log\left(2\sinh\left(\frac{t}{2}\right)\right)+i\pi\right]+2\left(\log\left(\alpha\sinh\left(\frac{t}{2}\right)\text{csch}\left(\frac{\alpha t}{2}\right)\right)+1\right)\Big).\label{eq:ExactResult}
\end{aligned}
\end{equation}
where $B(x,\beta,0)=\frac{x^{\beta}{}_{2}F_{1}(1,\beta,1+\beta,x)}{\beta}$
is the incomplete Beta function, $z\equiv1-e^{-t}$, $\alpha=\sqrt{1-24\eta_H}$, $H_{n}$ is the
harmonic function. 

 At order $1/c^2$, we calculated the order $\eta_H$ and $\eta_H^2$ terms in the expansion of the vacuum block in the parameter $\eta_H=\frac{h_H}{c}$, and at order $1/c^3$, we calculated the linear $\eta_H$ terms.

At order $1/c^2$,
\[
\log\CV\supset \frac{h_L}{c^2}\sum_{k=0}^\infty\eta_H^{k+1}\left(f_{20k}+h_Lf_{11k}+h_L^2f_{02k}\right).
\] 
The linear $\eta_H$ terms are $f_{200}$, $f_{110}=f_{101}$ and $f_{020}=f_{002}$. This first term is given in equation (\ref{eq:f200}), while the last two terms can be obtained from the expansion of $f_{10}$ and $f_{00}$.
The $\eta_H^2$ terms are 
\begin{equation}
\begin{aligned}
f_{201}=&\frac{432 \left((z-1) z (15 z-46)+4 \pi ^2 (z ((z-2) z-10)+12)\right) \log ^2(1-z)}{z^3}\\
&+\frac{864 (z (z (z (5 z-44)+103)-96)+33) \log ^4(1-z)}{z^4}+\frac{10368 (z-2)^2 \text{Li}_2(z){}^2}{z^2}\\
&+\frac{864 \left((9 z-46) (z-1)^2+(4 z (15-2 (z-2) z)-72) \log (z)\right) \log ^3(1-z)}{z^3}\\
&+\frac{5184 \text{Li}_2(z) \log (1-z) (z (z (7 z-32)+32)+2 (z-9) (z-2) (z-1) \log (1-z))}{z^3}\\
&+\frac{5184 \text{Li}_3(1-z) (z (z (5 z-14)+16)-4 (z-3) (z-1) (z+2) \log (1-z))}{z^3}\\
&+\frac{10368 \text{Li}_3(z) ((z-2) z+2 ((z-8) z+8) \log (1-z))}{z^2}+\frac{20736 (z-2) \text{Li}_4(1-z)}{z}\\
&+\frac{20736 ((z-6) z+6) \left(\text{Li}_4\left(\frac{z}{z-1}\right)+ \text{Li}_4(z)\right)}{z^2}+\frac{12960 (z-2) \text{Li}_2(z)}{z}\\
&+\frac{216 \left((z-2) z^2+96 (6-5 z) \zeta (3)-4 \pi ^2 (z (5 z-14)+16) z\right) \log (1-z)}{z^3}\\
&-\frac{144 \left(-525 z^2+180 (z (5 z-14)+16) \zeta (3)+8 \pi ^4 (z-2) z\right)}{5 z^2}\\
&+\frac{2592 (z (5 z-14)+16) \log (z) \log ^2(1-z)}{z^2},
\end{aligned}
\end{equation}
\begin{equation}
\begin{aligned}
f_{111}=&\frac{864 \left(3 z \left(z^2-8 z+7\right)+8 \pi ^2 \left(2 z^2-9 z+8\right)\right) \log ^2(1-z)}{z^3}+\frac{5184 (z-2) \text{Li}_2(z)}{z}\\
&+\frac{1728 (z (z (z (3 z-44)+127)-136)+51) \log ^4(1-z)}{z^4}-\frac{41472 (z-1) \text{Li}_2(z){}^2}{z^2}\\
&+\frac{3456 ((z-1) ((z-17) z+21)-6 (z (2 z-9)+8) \log (z)) \log ^3(1-z)}{z^3}\\
&+\frac{10368 ((z-7) z+7) \log (z) \log ^2(1-z)}{z^2}-\frac{41472 ((z-6) z+6) \text{Li}_3(z) \log (1-z)}{z^2}\\
&+\frac{20736 \text{Li}_2(z) \log (1-z) (z ((z-9) z+9)+(z ((z-14) z+34)-22) \log (1-z))}{z^3}\\
&+\frac{20736 \text{Li}_3(1-z) (z ((z-7) z+7)-2 (z (2 z-9)+8) \log (1-z))}{z^3}\\
&+\frac{41472 ((z-6) z+6) \left(\text{Li}_4(z)+\text{Li}_4\left(\frac{z}{z-1}\right)\right)}{z^2}+20736 \left(2-\frac{((z-7) z+7) \zeta (3)}{z^2}\right)\\
&+\frac{432 \left(3 (z-2) z^2+96 (z (2 z-9)+8) \zeta (3)-8 \pi ^2 ((z-7) z+7) z\right) \log (1-z)}{z^3},
\end{aligned}
\end{equation}
and $f_{021}=f_{012}$ can be obtained from the expansion of $f_{01}$.

At order $1/c^3$,   
\[
\log\CV\supset \frac{h_L}{c^3}\sum_{k=0}^\infty\eta_H^{k+1}\left(f_{30k}+h_Lf_{21k}+h_L^2f_{12k}+h_L^3f_{03k}\right),
\] 
The linear $\eta_H$ terms are 
\begin{equation}
\begin{aligned}
f_{300}=&\frac{864(2z(z(z^2-8z+17)-14)+9)\log^4(1-z)}{z^4}-\frac{20736 (z-1) \text{Li}_2(z){}^2}{z^2}\\
&+\frac{216 \log ^2(1-z) \left(8 \pi ^2 (z-2) \left(z^2-2\right)-108 z \left(z^2-1\right) \log (z)+73 z (z-1)^2\right)}{z^3}\\
&-\frac{864 (z-2) \log ^3(1-z) \left(4 \left(2 z^2-3\right) \log (z)-9 (z-1)^2\right)}{z^3}\\
&+\frac{432 \text{Li}_2(z) \left(73 (z-2) z^2+24 (z-1) \log (1-z) ((6-4 z) \log (1-z)-9 z)\right)}{z^3}\\
&+\frac{5184 \left(z^2-1\right) \text{Li}_3(1-z) (9 z+4 (z-2) \log (1-z))}{z^3}-\frac{20736 (2 z-3) \text{Li}_4\left(\frac{z}{z-1}\right)}{z^2}\\
&-\frac{5184 (z-2) \text{Li}_3(z) (9 z+4 (z-2) \log (1-z))}{z^2}+\frac{20736 (z-3) (z-1) \text{Li}_4(z)}{z^2}\\
&+\frac{20736 (z-2) \text{Li}_4(1-z)}{z}+\frac{192 \left(1215 \left(z^2-1\right) \zeta (3)+320 z^2-6 \pi ^4 (z-2) z\right)}{5 z^2}\\
&+\frac{12 \left((z-2) \left(z^2-1728 \zeta (3)\right)+648 \pi ^2 z \left(z^2-1\right)\right) \log (1-z)}{z^3},
\end{aligned}
\end{equation}
$f_{210}=f_{201}$, and $f_{120}=f_{102}$, $f_{030}=f_{003}$ can be obtained from the expansions of $f_{10}$ and $f_{00}$, respectively.

The terms $f_{200}$,$f_{201}$,$f_{111}$ and $f_{300}$ were derived for the first time in this work. We've checked these expressions against a direct small $z$ expansion up to $\CO(z^9)$ using the methods of \cite{Zamolodchikov:1987}. We've also analytically continue these results to the second sheet and checked that the they do contain the first few terms of (\ref{eq:expandedexp}) and (\ref{eq:1overcz}). Under this analytic continuation, the various logarithms and polylogarithms have monodromies
\begin{equation}
\begin{aligned}
	\log(1-z)& \rightarrow \log(1-z)-2\pi i,\\
	\text{Li}_n(z)&\rightarrow\text{Li}_n(z)+\frac{2\pi i}{(n-1)!}\log^{n-1}(z),\\
	\text{Li}_n(1-z)&\rightarrow\text{Li}_n(1-z),\\
	\text{Li}_n\left(\frac{z}{z-1}\right)&\rightarrow \text{Li}_n\left(\frac{z}{z-1}\right)-\frac{2\pi i}{(n-1)!}\log^{n-1}\left(\frac{z}{z-1}\right),	
\end{aligned}
\end{equation}
which can be derived from $\text{Li}_n(z)=\int_0^z\frac{\text{Li}_{n-1}(t)}{t}dt$ and $\text{Li}_1(z)=-\log(1-z)$.

\section{Direct Derivation of Leading Logs in the Lorentzian Regime}
\label{app:DirectLogResummation}

In subsection \ref{sec:logresum}, we presented a proof that the ``leading logs'' in the Lorentzian regime resum to form a correction to the leading singularity $(c z)^{-1}$ that appears at $\CO(1/c)$ in a large $c$ expansion. The proof given was somewhat indirect, however, and in this appendix we will give another proof that is more cumbersome, but more directly connected to the structure of the differential equations for the degenerate operators that are used order-by-order in $1/c$ in the rest of the paper.   In this appendix, for convenience we define
\be
\tilde{\CV}(z) = z^{2h_L} \CV(z),
\ee
so that for the vacuum block, $\tilde{\CV}(z) \stackrel{z\rightarrow 0}{\rightarrow} 1$ on the first sheet. 
 
From equation (\ref{eq:largecnull1}), at large $c$ the null equation of motion for the degenerate operator $\CO_{1,s}$ takes the form  
\be
(L_{-1}^{s}+ \CO(1/c)) \CO_{1,s} &=& 0 .
\label{eq:nullLs}
\ee
 In terms of $\tilde{\CV}$, (\ref{eq:nullLs})  translates into the differential equation 
\be
(\partial_z^s + \CO(1/c)) \left( z^{s-1} \tilde{\CV}(z) \right) &=& 0.
\ee
We will organize the solution to (\ref{eq:largecnull2}) in a series expansion of $\tilde{\CV}$:
\be
\tilde{\CV}(z) &\equiv& \tilde{\CV}_0(z) + \frac{1}{c} \tilde{\CV}_1(z) + \frac{1}{c^2} \tilde{\CV}_2(z) + \dots.
\ee
The lowest-order term then obey the following differential equation
\be
\partial_z^s (z^{s-1} \tilde{\CV}_0(z))&=& 0,
\label{eq:leadingnulleom}
\ee
whose general solution takes the form
\be
\tilde{\CV}_0(z) &=& \sum_{i=0}^{s-1} \frac{c_i}{z^i}
\label{eq:NullHomogeneous}
\ee
with $s$ free coefficients $c_i$. 
Of course, the relevant solution for the vacuum at $c\rightarrow \infty$ is $c_0=1, c_{i\ne 0} =0$.  But equally importantly, when we work to higher orders, all the solutions above will continue to be homogeneous solutions, and there will also be one particular solution at each order that arises because of the ``source'' from the lower order terms.  

There is a drastic simplification that occurs if we are 
interested only in the leading log terms. First, notice that none of the homogeneous solutions (\ref{eq:NullHomogeneous}) have logarithms in them. As a result, logarithms can be produced only by the ``particular'' solutions, which are integrals of the lower-order solutions. More precisely, the leading logarithms arise from integrating the lowest order solution and never introducing any ``homogeneous'' terms, since doing so would reduce the power of the logarithm.  Therefore, we can simply perform our analysis directly on the second sheet (where the differential equation must still be satisfied), and the unknown integration constants that enter at each step will not contaminate the leading logs.  

Using the expression (\ref{eq:leadinglight}), it is straightforward to extend this argument to leading logs in the heavy-light limit as well.  At infinite $c$, the general solution to (\ref{eq:leadinglight}) is
\be
\tilde{\CV}(t) &=& e^{\frac{1-s}{2}t} \sum_{j=0}^{s-1} c_j \exp \left[ t\left(2j-\frac{s-1}{2}\right)  \sqrt{1-24 \eta_H}\right].
\ee
These are all exponentials in $t$, i.e. powers in $(1-z)$.  Therefore, logarithms of $z$ can arise only from integrating source terms that are generated from the solution at lower orders.

Let us see how this works in practice, and along the way we will illustrate some points.  For simplicity, we will begin by solving for the leading logarithms in the conformal block $\< \CO_H \CO_H \CO_{1,2} \CO_{1,2}\>$.  Once we have gone through this case, it will be easy to see how to generalize to arbitrary degenerate operators. 

The exact equation of motion for $\tilde{\CV}(z)$ is
\begin{equation}
0 = (z-1) \left(\left(-4 (z-2) h_{1,2}+4 z-2\right) \tilde{\CV}'(z)+3 (z-1) z \tilde{\CV}''(z)\right)-2 z \left(2 h_{1,2}+1\right) \tilde{\CV}(z) h_H
\end{equation}
This can be solved in closed form by a hypergeometric function, but to illustrate our points we will solve it in a $1/c$ expansion.  At leading order it is just (\ref{eq:leadingnulleom}) with $s=2$.  At next order, it is
\be
\partial_z^2 (z \tilde{\CV}_1(z)) &=& -\frac{6 h_H z}{(1-z)^2},
\ee
which is easily solved:
\be
\tilde{\CV}_1(z) &=& c_0 + \frac{c_1}{z}  -\frac{6 (z-2) h_H \log (1-z)}{z}
\label{eq:V1piece}
\ee
We fix $c_0$ and $c_1$ on the first sheet by demanding that $\tilde{\CV}_1$ have the correct behavior (i.e., have leading term $\propto z$ in a small $z$ expansion), and then analytically continuing to the second sheet and taking small $z$ to find the small $z$ behavior on the second sheet.  Doing this, we find $c_0= 12 h_H, c_1=0$ on the first sheet.  Analytically continuing, this means that on the second sheet, 
\be
c_1=24 i \pi h_H, \qquad c_0 = -12 i \pi h_H.
\ee
Note that at this order, there are no logarithms $\log(z)$ in a small $z$ expansion, even on the second sheet:
\be
\tilde{\CV}_1(z) &=&  \frac{c_1}{z} + (c_0 - 12 h_H) + \CO(z).
\ee
To see the emergence of logarithms, we have to work to the next order in  $1/c$.  The equation of motion for $\tilde{\CV}_2$ is
\be
\partial_z^2 (z \tilde{\CV}_2(z)) &=&  -\frac{6 h_H z}{(1-z)^2} -\frac{6 \left(z \tilde{\CV}_1(z) h_H+(z-2) (z-1) \tilde{\CV}_1'(z)\right)}{(z-1)^2}
\ee
This can also be solved in closed form.  It again has two free parameters corresponding to the two homogeneous solutions, which we can fix the same way we fixed them for the free parameters in $\tilde{\CV}_1$.  However, we can instead apply an argument that will easily generalize to all higher orders, which is to expand the above equation of motion at small $z$ {\it directly on the second sheet}: 
\be
\partial_z^2 (z \tilde{\CV}_2(z)) &=& 6 c_1 \left( (1-h_H) + \frac{1}{z} + \frac{2}{z^2} + \CO(z) \right).
\ee
The solution to the above equation of motion is again easily determined:
\be
\tilde{\CV}_2(z) &=& 6 c_1 \left( -2\frac{\log(z)}{z} + \log(z) + \CO(z) \right) + \frac{d_1}{z} + d_0.
\ee
We do not need to determine the integration constants $d_0, d_1$, because they do not contaminate the leading logs!  Since the integration constants are always coefficients of the homogeneous solutions, this feature manifestly continues to all higher orders as well.
%

The above explicit demonstration was specific to the $\CO_{1,2}$ block, but it is straightforward to generalize to general degenerate operators.  
For all degenerate operators $\CO_{1,s}$, the $1/c$ piece $\tilde{\CV}_1$ is the same universal function (\ref{eq:V1piece}) (in fact, it is just the global conformal block for the stress tensor), with a coefficient that is linear in $h_{s,1}$:
\be
\tilde{\CV}_{1,s} &=& \frac{2h_H h_{1,s}}{c} z^2 {}_2F_1(2,2,4,z).
\ee 
We do not have to appeal to our knowledge that this is the stress tensor conformal block; (\ref{eq:V1piece}) is a derivation of $\tilde{\CV}_{s,1}$  since we know $\tilde{\CV}_{1,s} = \left( \lim_{c\rightarrow \infty} \frac{h_{1,s}}{h_{1,2}} \right)\tilde{\CV}_{1,2}$.  This means that generally, on the second sheet we have $\tilde{\CV}_1$ is given by
\be
c_1^{(s)} &=&\left( \lim_{c\rightarrow \infty} \frac{h_{1,s}}{h_{1,2}} \right)24 i \pi h_H = 24 i \pi h_H (s-1), \nn\\
c_0^{(s)} &=&\left( \lim_{c\rightarrow \infty} \frac{h_{1,s}}{h_{1,2}} \right) (-12 i \pi h_H) = 12 i \pi h_H (s-1).
\ee 
Thus, $\tilde{\CV}_2$ is generally given on the second sheet at small $z$ by 
\be
\partial_z^s \left( z^{s-1} \tilde{\CV}_2 \right)
&=&  12  A_s \frac{z \tilde{\CV}_1(z)}{z^2} + \CO(1/z) \nn\\
&=& \frac{12 c_1^{(s)}A_s}{z^2} + \CO(1/z),
\label{eq:nulleomF2}
\ee
(where $A_s$ depends on $s$ but will be determined momentarily).  We have taken advantage of the fact that $z \tilde{\CV}_1(z)$ is regular at $z\rightarrow 0$ since $c_1/z$ was the most singular term generated at this order, and so by scaling $z \tilde{\CV}_1(z)/z^2$ is the most singular term generated in the null equation of motion above. 
The solution to (\ref{eq:nulleomF2}) is clearly
\be
\tilde{\CV}_2(z) &=& 12 c_1^{(s)} \frac{\log(z)}{z} \frac{A_s}{(s-2)!}+ \CO(\log(z)).
\ee
We can easily fix $A_s$ since $\tilde{\CV}_2$ is completely determined for any $h_{1,s}$ by just the two function $\tilde{\CV}_{1,2}$ and $\tilde{\CV}_{2,2}$; therefore once we calculate $\tilde{\CV}_2$ for two values of $s$, we know it for all $s$.  A simple computation shows that $A_2=A_3=1$.  Demanding consistency of the above equation with all $r$ immediately fixes
\be
\frac{A_s}{(s-2)!}=1.
\ee

Finally, to get the leading logs, we can just iterate at higher orders, since the only way to get double logs is to integrate single logs (which first appear in $\tilde{\CV}_2$), and the only way to get triple logs is to integrate double logs (which first appear in $\tilde{\CV}_3$), etc.  So for instance, in the equation of motion for $\tilde{\CV}_3$, we can just look at $\tilde{\CV}_2$ in the source terms, since this is the only contribution that has a single log.  But the relation between $\tilde{\CV}_3$ and $\tilde{\CV}_2$ at leading order in $1/c$ is the same as the relation between $\tilde{\CV}_2$ and $\tilde{\CV}_1$ at leading order in $1/c$:
\be
\partial_z^s \left( z^{s-1} \tilde{\CV}_3 \right) &=& 12  (s-2)! \frac{z \tilde{\CV}_2(z)}{z^2} + \CO(1/z),
\ee
and so on.
Keeping track of just the most singular leading log terms, we see that at each order 
\begin{equation}
\tilde{\CV}_{n}(z) \supset 12c_1 \frac{\log^{n-1}(z)}{z (n-1)!} + \CO(z^0,\log(z)) \rightarrow \tilde{\CV}_{n+1}(z) \supset -12 c_1 \frac{\log^n(z)}{z n!} + \CO(z^0, \log(z)),
\end{equation}
which proves that the leading singularity in the leading logs exactly exponentiates to all orders. 



\section{Leading Contribution to the Vacuum blocks in $\mathcal{N}=1$  SCFTs} \label{appendixn=1}
In this section, we are going to prove that the heavy-light vacuum block $\CV_{\phi_L\phi_L\phi_H\phi_H}$ in $\mathcal{N}=1$ $\text{SCFT}$s is the same as the vacuum block in non-supersymmetric $\text{CFT}$s at leading order of the large $c$ limit, meaning that it only gets contributions from the pure Virasoro generators at this order.

The commutators of the symmetry generators with the component fields of a superfield  $\Phi\left(Z\right)=\phi_h\left(z\right)+\theta\psi_{h+\frac{1}{2}}\left(z\right)$ are
\begin{equation}\label{eq:n=1commutator}
\begin{aligned}
\left[L_{n},\phi\left(z\right)\right] &= z^{n}\left[h\left(n+1\right)+z\partial_{z}\right]\phi,\\
\left[L_{n},\psi\left(z\right)\right] &= z^{n}[(h+\frac{1}{2})\left(n+1\right)+z\partial_{z}]\psi,\\
\left[G_{r},\phi\left(z\right)\right] &= z^{r+\frac{1}{2}}\psi,\\
\left\{G_r,\psi(z)\right\} &= z^{r-\frac{1}{2}}\left[h\left(2r+1\right)+z\partial_{z}\right]\phi, \qquad n\in \mathbb{Z};\ r\in \mathbb{Z}+\frac{1}{2}.
\end{aligned}
\end{equation}

The vacuum block $\CV_{\phi_L\phi_L\phi_H\phi_H}$ is the contribution to $\left\langle\phi_H(\infty)\phi_H(1)\phi_L(z)\phi_L(0)\right\rangle$ from an irreducible representation of the super-Virasoro algebra whose highest weight state is the vacuum $\left|0\right>$. The vacuum state is annihilated by $L_n$ and $G_r$ for $n\ge -1$ and  $r\ge -\frac{1}{2}$. Besides the vacuum state, other states in this representation are the descendants of the vacuum, which can be obtained by acting on the vacuum with $L_{-n}$ and $G_{-r}$ for $n\ge2$ and $r\ge\frac{3}{2}$.
To get the vacuum block, we can insert a projection operator into the four-point function:
\begin{equation}
\CV_{\phi_L\phi_L\phi_H\phi_H}=\frac{\left\langle\phi_H(\infty)\phi_H(1)\CP_0\phi_L(z)\phi_L(0)\right\rangle}{\left\langle\phi_H(\infty)\phi_H(1)\right\rangle}.
\end{equation}
At leading order of the large $c$ limit, we can use the approximate projection operator
\be
\CP_0\approx\sum_{\{n_i, r_j\}}\frac{\left. G_{-r_{j}}\cdots G_{-r_{1}}L_{-n_{i}}\cdots L_{-n_{1}}|0\right\rangle \left\langle 0|L_{n_{1}}\cdots L_{n_{i}}G_{r_{1}}\cdots G_{r_{j}}\right.}{{\left\langle L_{n_{1}}\cdots L_{n_{i}}G_{r_{1}}\cdots G_{r_{j}}G_{-r_{j}}\cdots G_{-r_{1}}L_{-n_{i}}\cdots L_{-n_{1}}\right\rangle }}. \label{eq:projection}
\ee
with $n_i\in \mathbb{Z}$ and $r_j\in \mathbb{Z}+\frac{1}{2}$, because the states $G_{-r_{j}}\cdots G_{-r_{1}}L_{-n_{i}}\cdots L_{-n_{1}}\left|0\right\rangle $ are orthogonal with each other at this order \footnote{The proof is similar to that for non-susy CFTs with only Virasoro generators, see appendix B of \cite{Fitzpatrick:2014vua}.}. We can arrange the order of the generators such that  $n_i\ge\cdots \ge n_1\ge2$ and $r_j\ge \cdots \ge r_1\ge \frac{3}{2}$. Denote the level of each state as $N+R$, where $N=\sum_{l=1}^i n_l$ and $R=\sum_{l=1}^j r_l$. Notice that in the above equation (\ref{eq:projection}), at each level $N+R$, we should only sum over independent states. For example, at level 3, we only have $L_{-3}$, because $G_{-\frac{3}{2}}G_{-\frac{3}{2}}=L_{-3}$ and shouldn't be included.

Consider a state $G_{-r_{j}}\cdots G_{-r_{1}}L_{-n_{i}}\cdots L_{-n_{1}}\left|0\right\rangle $, its contribution to $\CV_{\phi_L\phi_L\phi_H\phi_H}$ is
\begin{equation}\label{eq:n=1direct}
\frac{\left\langle \phi_{H}\left(\infty\right)\phi_{H}\left(1\right)G_{-r_{j}}\cdots G_{-r_{1}}L_{-n_{i}}\cdots L_{-n_{1}}|0\right\rangle  \left\langle 0|L_{n_{1}}\cdots L_{n_{i}}G_{r_{1}}\cdots G_{r_{j}}\phi_{L}\left(z\right)\phi_{L}\left(0\right)\right\rangle }{\left\langle\phi_H(\infty)\phi_H(1)\right\rangle\left\langle L_{n_{1}}\cdots L_{n_{i}}G_{r_{1}}\cdots G_{r_{j}}G_{-r_{j}}\cdots G_{-r_{1}}L_{-n_{i}}\cdots L_{-n_{1}}\right\rangle} 
\end{equation}
In the large c limit, the normalization factor in the  denominator scales as 
\[
\left\langle L_{n_{1}}\cdots L_{n_{i}}G_{r_{1}}\cdots G_{r_{j}}G_{-r_{j}}\cdots G_{-r_{1}}L_{-n_{i}}\cdots L_{-n_{1}}\right\rangle \sim c^{N+R}
\]
because the commutation of each pair of generators $G_r$ with $G_{-r}$ or $L_{n}$ with $L_{-n}$ will give us one power of $c$ (\ref{eq:n=1commutation}). In the numerator, $\left\langle 0|L_{n_{1}}\cdots L_{n_{i}}G_{r_{1}}\cdots G_{r_{j}}\phi_{L}\left(z\right)\phi_{L}\left(0\right)\right\rangle $ is order $O(1)$, because the commutation of these generators with $\phi_L$ will not give us $c$ or $h_H$. And the remaining part in (\ref{eq:n=1direct}) scales as 
\begin{equation}
\frac{\left\langle \phi_{H}\left(\infty\right)\phi_{H}\left(1\right)G_{-r_{j}}\cdots G_{-r_{1}}L_{-n_{i}}\cdots L_{-n_{1}}|0\right\rangle}{\left\langle\phi_H(\infty)\phi_H(1)\right\rangle} \sim h_{H}^{N+R/2}.
\end{equation}
The reason is that when we commute one $L_{-n}$ with $\phi_H$ we'll get one power of $h_H$, but we need to commute two $G_{-r}$s with $\phi_H$ to get one power of $h_H$ as can be seen from the commutation relations (\ref{eq:n=1commutator}). So in the heavy-light limit, with $\eta_H=\frac{h_H}{c}$ fixed, the contribution of (\ref{eq:n=1direct}) will be order $O(c^{-R/2})$. This means that  at order $c^{0}$ (that is, $R=0$), the heavy-light vacuum block  $\CV_{\phi_L\phi_L\phi_H\phi_H}$ in $\mathcal{N}=1$ SCFTs will only get contributions from the pure Virasoro generators, which make it the same as that in non-susy $\text{CFT}$s at leading order. This is also true for the vacuum blocks $\CV_{\psi_L\psi_L\phi_H\phi_H}$.

\section{Details of the $\mathcal{N}=2$ $\text{SCFT}$ Calculations}
\subsection{Superconformal Ward Identities}\label{AppendixWard}
$N$-point functions $F_N\equiv\left\langle\Phi_1(Z_1)\Phi_2(Z_2)\cdots\Phi_N(Z_N)\right\rangle$ in $\mathcal{N}=2$ SCFTs satisfy the following eight superconformal Ward identities
\begin{align}\label{eq:n=2Ward}
&L_{-1} : \sum_{i=1}^N \partial_{z_i}F_N=0,\nonumber\\
&L_0 : \sum_{i=1}^N (2z_i\partial_{z_i}+2h_i+\theta_i\partial_{\theta_i}+\overline{\theta}_i\partial_{\overline{\theta}_i})F_N=0,\nonumber\\
&L_{1}: \sum_{i=1}^N (z_i^2\partial_{z_i}+z_i(2h_i+\theta_i\partial_{\theta_i}+\overline{\theta}_i\partial_{\overline{\theta}_i})+q_i\theta_i\overline{\theta}_i)F_N=0,\nonumber\\
&J_0: \sum_{i=1}^N (\overline{\theta}_{i}\partial_{\overline{\theta}_i}-\theta_{i}\partial_{\theta_i}+q_i)F_N=0\\
&G_{-\frac{1}{2}}, \overline{G}_{-\frac{1}{2}} : \sum_{i=1}^N (\partial_{\overline{\theta}_i}-\theta_{i}\partial_{z_{i}})F_N=\sum_{i=1}^N (\partial_{\theta_i}-\overline{\theta}_{i}\partial_{z_{i}})F_N=0,\nonumber\\
&G_{\frac{1}{2}}: \sum_{i=1}^N[z_{i}(\partial_{\overline{\theta}_i}-\theta_{i}\partial_{z_{i}})-\theta_{i}(2h_i+q_i+\overline{\theta}_{i}\partial_{\overline{\theta}_i})]F_N=0,\nonumber\\
&\overline{G}_{\frac{1}{2}}:\sum_{i=1}^N[z_{i}(\partial_{\theta_i}-\overline{\theta}_{i}\partial_{z_{i}})-\overline{\theta}_{i}(2h_i-q_i+\theta_{i}\partial_{\theta_i})]F_N=0.\nonumber
\end{align}
Specifically, the three identities corresponding to $L_{-1}$, $G_{-\frac{1}{2}}$ and $\overline{G}_{-\frac{1}{2}}$ were used in the simplification of the super null-state equation (\ref{eq:n=2nullstateequation}).

\subsection{Leading Contributions to the Vacuum Blocks} \label{Appendixn=2Leading}
Similar to the reasoning of $\mathcal{N}=1$ (appendix \ref{appendixn=1}), the vacuum block $\CV_{\phi^{-q_L}_L\phi^{q_L}_L\phi^{-q_H}_H\phi^{q_H}_H}$ in $\mathcal{N}=2$ SCFTs will not get contribution from the generators $G_r$ and $\overline{G}_r$ at leading order of large $c$ limit, which makes it the same as the vacuum block of a theory with only Virasoro and $U(1)$ symmetry at this order. This can be seen from the commutation relations of these symmetry generators with the lowest component field $\phi(z)$ of a superfield $
\Phi_{h}^{q}\left(Z\right)=\phi_{h}^{q}\left(z\right)+\theta\overline{\psi}_{h+\frac{1}{2}}^{q-1}\left(z\right)+\overline{\theta}\psi_{h+\frac{1}{2}}^{q+1}\left(z\right)+\theta\overline{\theta}\lambda_{h+1}^{q}\left(z\right).$ For simplicity, in the following subsections, we only keep the superscripts and subscripts when necessary.

Commutation relations of the generators with the component field $\phi(z)$ are 
\begin{equation}\label{eq:n=2PhiCommutation}
\begin{aligned}
\left[L_{n},\phi\left(z\right)\right]=&z^n\left[\left(n+1\right)h+z\partial_{z}\right]\phi, \quad 
\left[J_{n},\phi\left(z\right)\right]=qz^{n}\phi, \\
\left[G_{r},\phi\left(z\right)\right]=&z^{r+\frac{1}{2}}\psi, \quad 
\left[\overline{G}_{r},\phi\left(z\right)\right]=z^{r+\frac{1}{2}}\overline{\psi}.
\end{aligned}
\end{equation}
The last two commutators are exactly the same as that of the fermionic generator $G_r$ with $\phi$ in $\mathcal{N}=1$ SCFTs (\ref{eq:n=1commutator}), which upon the same reasoning means that when summing over descendant states of the vacuum to get $\CV_{\phi^{-q_L}_L\phi^{q_L}_L\phi^{-q_H}_H\phi^{q_H}_H}$, those states having $G_r$ or $\overline{G}_r$ in them will not contribute at leading order of the large $c$ limit. We can also easily see that some other vacuum blocks, such as $\CV_{\lambda_L^{-q_L}\phi_L^{q_L}\phi_H^{-q_H}\phi_H^{q_H}}$,   $\CV_{\phi_L^{-q_L}\lambda_L^{q_L}\phi_H^{-q_H}\phi_H^{q_H}}$ and $\CV_{\lambda_L^{-q_L}\lambda_L^{q_L}\phi_H^{-q_H}\phi_H^{q_H}}$, also only get contributions from Virasoro and $U(1)$ generators. This point will be used in the calculation of subsection \ref{AppendixLambdaDecom}. Note that to construct the projection operator for $\mathcal{N}=2$, the Hermiticity conditions among these generators are $ L_n^\dagger=L_{-n}$, $J_n^\dagger=J_{-n}$, $G_r^\dagger=\overline{G}_{-r}$, $\overline{G}_{r}^\dagger=G_{-r}$. And the vacuum $\left|0\right\rangle$ in $\mathcal{N}=2$ is annihilated by $L_n,J_m,G_{r},\overline{G}_r$ for $n\ge-1,m\ge0,r\ge-\frac{1}{2}$.

For completeness, the commutation relations of other component fields are 
\begin{align}\label{eq:n=2FieldCommuations}
\left[L_{n},\psi\left(z\right)\right]&=z^{n}[(h+\frac{1}{2})\left(n+1\right)+z\partial_{z}]\psi,\nonumber\\
\left[L_{n},\overline{\psi}\left(z\right)\right]&=z^{n}[(h+\frac{1}{2})\left(n+1\right)+z\partial_{z}]\overline{\psi},\nonumber\\
\left[L_{n},\lambda\left(z\right)\right]&=z^n\left[\left(h+1\right)\left(n+1\right)z+z\partial_z\right]\lambda+\frac{1}{2}n(n+1)qz^{n-1}\phi,\nonumber\\
\left\{ G_{r},\psi\left(z\right)\right\}&=\left\{ \overline{G}_{r},\overline{\psi}(z)\right\}=0\nonumber\\
\left\{ G_{r},\overline{\psi}\left(z\right)\right\} &=z^{r-\frac{1}{2}}[(r+\frac{1}{2})\left(2h+q\right)+z\partial_z]\phi+z^{r+\frac{1}{2}}\lambda,\nonumber\\
\left\{ \overline{G}_{r},\psi\left(z\right)\right\} & =z^{r-\frac{1}{2}}[(r+\frac{1}{2})\left(2h-q\right)+z\partial_z]\phi-z^{r+\frac{1}{2}}\lambda,\nonumber\\
\left[G_{r},\lambda\left(z\right)\right]&=-z^{r-\frac{1}{2}}[(r+\frac{1}{2})\left(2h+q+1\right)+z\partial_{z}]\psi,\\
\left[\overline{G}_{r},\lambda\left(z\right)\right]&=z^{r-\frac{1}{2}}[(r+\frac{1}{2})\left(2h-q+1\right)+z\partial_{z}]\overline{\psi},\nonumber\\
\left[J_{n},\psi\left(z\right)\right]&=\left(q+1\right)z^{n}\psi, \nonumber\\
\left[J_{n},\overline{\psi}\left(z\right)\right]&=\left(q-1\right)z^{n}\overline{\psi}\qquad\nonumber\\
\left[J_{n},\lambda\left(z\right)\right]&=qz^n \lambda +2hnz^{n-1}\phi.\nonumber
\end{align}

\subsection{Correlation Functions with Descendant Component Fields} \label{appendixn=2descendant}
In this subsection, we are going to derive the relationships between correlation functions with descendant fields and correlation functions with only primary fields. These relationships are also true for the corresponding vacuum blocks. Specifically, we only consider the lowest component primary field $\phi^q_h$ and its descendants that are relevant to our calculation. 

For correlation functions involving $\left(L_{-1}\phi\right)\left(z\right)$,
since $\left(L_{-1}\phi\right)\left(z\right)=\partial_{z}\phi\left(z\right)$,
we have 
\be
\left\langle \left(L_{-1}\phi\right)\left(z\right)X\right\rangle =\partial_{z}\left\langle \phi(z) X\right\rangle 
\ee
where $X$ is an assembly of primary or descendant component fields. If there
are more than one $\left(L_{-1}\phi\right)$, we just need to take
the derivatives in succession with respect to the coordinate of each
$\left(L_{-1}\phi\right)$. 

For correlation functions involving only one descendant $\left(J_{-n}\phi\right)$, we have 
\begin{equation}
\begin{aligned}
\left\langle \left(J_{-n}\phi\right)\left(z_{1}\right)Y\right\rangle& =\frac{1}{2\pi i}\oint_{z_1}dz\left(z-z_{1}\right)^{-n}\left\langle J\left(z\right)\phi\left(z_{1}\right)Y\right\rangle \\
& =-\frac{1}{2\pi i}\sum_{i=2}^{N}\oint_{z_{i}}dz\left(z-z_{1}\right)^{-n}\frac{q_{i}\left\langle \phi\left(z_{1}\right)Y\right\rangle }{z-z_{i}}\label{eq:CorrelatorwithJ-n}\\
& =-\sum_{i=2}^{N}\frac{q_{i}\left\langle \phi\left(z_{1}\right)Y\right\rangle }{\left(z_{i}-z_{1}\right)^n}
\end{aligned}
\end{equation}
where $Y=\phi_{2}\left(z_{2}\right)\cdots\phi_{N}\left(z_{N}\right)$
is an assembly of primary fields with conformal dimensions $h_{i}$
and $U\left(1\right)$ charge $q_{i}$, and we have used the OPE $J(z)\phi_i(z_i)\sim \frac{q_i\phi_i(z_i)}{z-z_i}$ in the second line.

For correlation functions involving two $\left(J_{-n}\phi\right)$s, we need to know the OPE $J\left(z\right)\left(J_{-n}\phi\right)\left(w\right)$, which can be written as 
\begin{equation}
J(z)(J_{-n}\phi)(w)=\sum_{k>0}\frac{(J_{k,-n}\phi)(w)}{(z-w)^{k+1}}+\sum_{k\ge 0} \frac{(J_{-k,-n}\phi)(w)}{(z-w)^{1-k}}
\end{equation}
In the first sum, since $[J_k,J_{-n}]=\frac{c}{3}k\delta_{k-n,0}$ and $(J_k\phi)(w)=0$ for $k>0$, only the term with $k=n$ is non-zero. In the second sum, only the term with $k=0$ is singular.  So we have 
\begin{equation}
\begin{aligned}
J(z)(J_{-n}\phi)(w)\sim&\frac{(J_{n,-n}\phi)(w)}{(z-w)^{n+1}}+ \frac{(J_{0,-n}\phi)(w)}{z-w}\\
\sim &\frac{nc}{3}\frac{\phi(w)}{(z-w)^{n+1}}+\frac{q(J_{-n}\phi)(w)}{z-w}
\end{aligned}
\end{equation}
where $q$ is the $U(1)$ charge of $\phi$ (and $J_{-n}$ will not change the $U(1)$ charge) and $\sim$ means that in the RHS we omit terms that are regular. For $n=1$, the OPE of $J\left(z\right)$ with $\left(J_{-1}\phi\right)\left(w\right)$ is 
\begin{equation}
J\left(z\right)(J_{-1}\phi)\left(w\right)\sim\frac{c}{3}\frac{\phi\left(w\right)}{\left(z-w\right)^{2}}+\frac{q\left(J_{-1}\phi\right)\left(w\right)}{\left(z-w\right)}
\end{equation}
In the calculation of this paper, we only need $\left\langle \left(J_{-1}\phi_{1}\right)\left(z_{1}\right)\left(J_{-1}\phi_{2}\right)\left(z_{2}\right)Y\right\rangle $
with $Y=\phi_{3}\left(z_{3}\right)\cdots\phi_{N}\left(z_{N}\right)$ an assembly of primary fields.
Using the above OPE, we have 
\begin{equation}
\begin{aligned}
\left\langle \left(J_{-1}\phi_{1}\right)\left(z_{1}\right)\left(J_{-1}\phi_{2}\right)\left(z_{2}\right)Y\right\rangle &=  \frac{1}{2\pi i}\oint_{z_{1}}dz\frac{\left\langle J\left(z\right)\phi_1\left(z_{1}\right)\left(J_{-1}\phi_{2}\right)\left(z_{2}\right)Y\right\rangle }{z-z_{1}}\\
&= -\frac{1}{2\pi i}\oint_{z_{2}}\frac{dz}{z-z_{1}}\left\langle \phi_{1}\left[\frac{c}{3}\frac{\phi_{2}}{\left(z-z_{2}\right)^{2}}+\frac{q_{2}\left(J_{-1}\phi_{2}\right)}{\left(z-z_{2}\right)}\right]Y\right\rangle \\&\quad -\frac{1}{2\pi i}\sum_{i=3}^{N}\oint_{z_{i}}\frac{dz}{z-z_1}\frac{q_{i}\left\langle \phi_{1}\left(J_{-1}\phi_{2}\right)Y\right\rangle }{z-z_{i}}\\
&=  \frac{c}{3}\frac{\left\langle \phi_{1}\phi_{2}Y\right\rangle }{\left(z_{2}-z_{1}\right)^{2}}-\sum_{i=2}^N\frac{q_{i}\left\langle \phi_{1}\left(J_{-1}\phi_{2}\right)Y\right\rangle }{z_{i}-z_{1}}\\
&= \frac{c}{3}\frac{\left\langle \phi_{1}\phi_{2}Y\right\rangle }{\left(z_{2}-z_{1}\right)^{2}}+ \sum_{i=2}^{N}\sum_{j=1,j\ne2}^{N}\frac{q_{i}q_{j}\left\langle \phi_{1}\phi_{2}Y\right\rangle }{\left(z_{i}-z_{1}\right)\left(z_{j}-z_{2}\right)}
\end{aligned}
\end{equation}
where in the second line, we used the OPE of $J(z)(J_{-1}\phi_2)(z_2)$ and $J(z)Y$ (or equation (\ref{eq:CorrelatorwithJ-n})), and in the last line, we used equation equation (\ref{eq:CorrelatorwithJ-n}).

\subsection{Decomposition of $\lambda^q_{h+1}$}\label{AppendixLambdaDecom}
In this subsection, we'll show that $\lambda_{h+1}^q$with conformal dimension $h+1$ and $U(1)$ charge $q$ can be written as 
\begin{equation}\label{eq:Appendixn=2LambdaDecompose}
\lambda_{h+1}^q(z)=\frac{12h^{2}-3q^{2}}{2ch-3q^{2}}(J_{-1}\phi_h^q)(z)+\frac{q(c-6h)}{2ch-3q^{2}}(L_{-1}\phi_h^q)(z)+\tilde{\lambda}_{h+1}^q(z),
\end{equation}
where $\tilde{\lambda}_{h+1}^q$ is a Virasoro and $U(1)$ primary with conformal dimension $h+1$ and $U(1)$ charge $q$, in the sense that $L_{0}\tilde{\lambda}^q_{h+1}=(h+1)\tilde{\lambda}^q_{h+1}$, $J_{0}\tilde{\lambda}^q_{h+1}=q\tilde{\lambda}^q_{h+1}$ and $L_{n}\tilde{\lambda}^q=J_{n}\tilde{\lambda}^q=0,n\ge1$. In the following calculation, for simplicity, we only keep  the superscripts and subscripts when necessary.

$\lambda^q$ can be obtained by acting on the lowest component field $\phi^q$ with $G_{-\frac{1}{2}}$ and $\overline{G}_{-\frac{1}{2}}$:
\begin{equation}
\lambda^q  =\frac{1}{2}\left(G_{-\frac{1}{2}}\overline{G}_{-\frac{1}{2}}-\overline{G}_{-\frac{1}{2}}G_{-\frac{1}{2}}\right)\phi^q =\left(L_{-1}-\overline{G}_{-\frac{1}{2}}G_{-\frac{1}{2}}\right)\phi^q.\label{eq:lambda1}
\end{equation}
Suppose $\lambda^q$ can be written as 
\begin{equation}
\lambda^q=AJ_{-1}\phi^q+BL_{-1}\phi^q+\tilde{\lambda}^q \label{eq:lambda2},
\end{equation}
where $A$ and $B$ are two constants depending on $h$ and $q$, and $\tilde{\lambda}^q$ is a Virasoro and $U(1)$ primary.
Acting on (\ref{eq:lambda1}) and (\ref{eq:lambda2}) with $L_{1}$
and $J_{1}$, we get two equations
\begin{align*}
L_{1}\left(L_{-1}-\overline{G}_{-\frac{1}{2}}G_{-\frac{1}{2}}\right)\phi^q &= L_{1}\left(AJ_{-1}\phi^q+BL_{-1}\phi^q+\tilde{\lambda}^q\right),\\
J_{1}\left(L_{-1}-\overline{G}_{-\frac{1}{2}}G_{-\frac{1}{2}}\right)\phi^q &= J_{1}\left(AJ_{-1}\phi^q+BL_{-1}\phi^q+\tilde{\lambda}^q\right).
\end{align*}
Using the commutation relation of these generators (\ref{eq:n=2commutation}), we have 
\begin{align*}
q\phi^q=&\left(Aq+2hB\right)\phi^q,\\
2h\phi^q=&\left(\frac{Ac}{3}+Bq\right)\phi^q.
\end{align*}
Solving these equations, we get 
\begin{equation}
A=\frac{12h^{2}-3q^{2}}{2ch-3q^{2}},B=\frac{q(c-6h)}{2ch-3q^{2}}.
\end{equation}
which give us the decomposition as equation (\ref{eq:Appendixn=2LambdaDecompose}). Note the $A$ is invariant but $B$ changes sign when $q$ is changed to $-q$, so the decomposition of $\lambda^{-q}$ is 
 \begin{equation}
 \lambda^{-q}=AJ_{-1}\phi^{-q}-BL_{-1}\phi^{-q}+\tilde{\lambda}^{-q} \label{eq:LambdaMinusq}.
 \end{equation}
 
The commutation of $\tilde{\lambda}^q$ with the Virasoro and $U(1)$ generators can  be derive from those of $\lambda^q$ (\ref{eq:n=2FieldCommuations}):
\begin{equation}
\begin{aligned}
 \left.\right.[L_{n},\tilde{\lambda}^q(z)] &=[L_{n},\lambda^q(z)]-A[L_{n},(J_{-1}\phi^q)(z)]-B[L_{n},(L_{-1}\phi^q)(z)],\\
[J_{n},\tilde{\lambda}^q(z)] &=[J_{n},\lambda^q(z)]-A[J_{n},(J_{-1}\phi^q)(z)]-B[J_{n},(L_{-1}\phi^q)(z)].
\end{aligned}
\end{equation}
The commutation relations of $J_{-1}\phi^q(z)$ on the RHS can be derive from the OPE of $T(w)$ and $J(w)$ with  $J_{-1}\phi^q(z)$:
\begin{equation}
\begin{aligned}
T(w)(J_{-1}\phi^q)(z)& \sim \frac{q\phi(z)}{(w-z)^3}+\frac{(h+1)(J_{-1}\phi)(z)}{(w-z)^2}+\frac{\partial_z (J_{-1}\phi)(z)}{w-z},\\
J\left(w\right)(J_{-1}\phi^q)\left(z\right)&\sim\frac{c}{3}\frac{\phi\left(z\right)}{\left(w-z\right)^{2}}+\frac{q\left(J_{-1}\phi\right)\left(z\right)}{\left(w-z\right)}.
\end{aligned}
\end{equation}
and the results are 
\begin{equation}
\begin{aligned}
\left.\right.[L_n,(J_{-1}\phi^q)(z)]&=z^{n-1}[\frac{1}{2}(n+1)nq\phi^q+(h+1)(n+1)z(J_{-1}\phi^q)+z^2\partial_z(J_{-1}\phi^q)]\\
[J_n,(J_{-1}\phi^q)(z)]&=z^{n-1}[\frac{c}{3}n\phi^q+zq(J_{-1}\phi^q)].
\end{aligned}
\end{equation}
The commutation relations of $L_n$ and $J_n$ with $L_{-1}\phi^q(z)=\partial_z \phi^q(z)$ are just the derivative of the commutation relations of $L_n$ and $J_n$ with $\phi^q(z)$ given in (\ref{eq:n=2PhiCommutation}):
\begin{equation}
\begin{aligned}
\left[L_{n},(L_{-1}\phi^q)\left(z\right)\right]&=z^{n-1}[n\left(n+1\right)h+\left(n+1\right)z\partial_{z}+z^2\partial_z^2]\phi^q,\\
\left[J_{n},(L_{-1}\phi^q)\left(z\right)\right]&=qz^{n-1}(n+z\partial_z)\phi^q.
\end{aligned}
\end{equation}

Putting everything together, we finally get
\begin{equation}
\left.\right.[L_{n},\tilde{\lambda}^q(z)]=z^n[\left(h+1\right)\left(n+1\right)+z\partial_{z}]\tilde{\lambda}^q,\quad
[J_{n},\tilde{\lambda}^q(z)]=qz^{n}\tilde{\lambda}^q.
\end{equation}
Comparing these two commutations with those for $\phi^q_h$ \eqref{eq:n=2PhiCommutation}, we can see that under the action of Virasoro and $U(1)$ generators, $\lambda^q_{h+1}$ acts like $\phi^q_h$ but with conformal dimension $h+1$.

To derive the normalization of two-point function $\langle \tilde{\lambda}^{-q}(z_1)\tilde{\lambda}^{q}(z_2)\rangle $, we need to use the two-point function of $\lambda^q$ and $\lambda^{-q}$, which can be read off from the two-point function of two superfields (\ref{eq:n=2twopt}):
\begin{equation}
\left\langle \lambda_{h+1}^{-q}\left(z_{1}\right)\lambda_{h+1}^q\left(z_{2}\right)\right\rangle =\frac{2h\left(2h+1\right)}{z_{21}^{2h+2}}.\label{eq:n=2lambdalambda}
\end{equation}
Substituting the decompositions of $\lambda_{h+1}^q$ and $\lambda_{h+1}^{-q}$ in the above
two-point function, we can express $\langle \tilde{\lambda}^{-q}(z_1)\tilde{\lambda}^{q}(z_2)\rangle$ as
\begin{align}\label{eq:n=2lambdatildelambdatilde}
\langle \tilde{\lambda}^{-q}(z_1)\tilde{\lambda}^{q}(z_2)\rangle =&\left\langle \lambda^{-q}\lambda^q\right\rangle -A^{2}\left\langle (J_{-1}\phi^{-q})(J_{-1}\phi^{q})\right\rangle +B^{2}\left\langle (L_{-1}\phi^{-q}) (L_{-1}\phi^q)\right\rangle \nonumber\\&-AB\left\langle (J_{-1}\phi^{-q}) (L_{-1}\phi^q)\right\rangle +AB\left\langle (L_{-1}\phi^{-q}) (J_{-1}\phi^q)\right\rangle.
\end{align}
The terms on the RHS are easy to calculate using the equations derived in last subsection \ref{appendixn=2descendant} and the two-point function $\left\langle \phi^{-q}(z_1)\phi^q(z_2))\right\rangle=\frac{1}{z_{21}^{2h}}$. The results are as follow
\begin{equation}
\begin{aligned}
\left\langle (J_{-1}\phi^{-q})(J_{-1}\phi^q)\right\rangle = & \frac{q^{2}+\frac{c}{3}}{z_{21}^{2h+2}},\\
\left\langle (L_{-1}\phi^{-q})(L_{-1}\phi^q)\right\rangle =&\partial_{z_{1}}\partial_{z_{2}}\frac{1}{z_{21}^{2h}}=\frac{-2h\left(2h+1\right)}{z_{21}^{2h+2}},\\
\left\langle (J_{-1}\phi^{-q})(L_{-1}\phi^q)\right\rangle =&\partial_{z_{2}}\frac{-q\left\langle \phi^{-q}\phi^q\right\rangle }{z_{21}}=\frac{\left(2h+1\right)q}{z_{21}^{2h+2}},\\
\left\langle (L_{-1}\phi^{-q})(J_{-1}\phi^{q})\right\rangle =&\partial_{z_{1}}\frac{q\left\langle \phi^{-q}\phi^q\right\rangle }{z_{12}}=\frac{-\left(2h+1\right)q}{z_{21}^{2h+2}}.
\end{aligned}
\end{equation}
Plugging these equations and equation (\ref{eq:n=2lambdalambda}) back in  equation (\ref{eq:n=2lambdatildelambdatilde}), we get 
\begin{equation}\label{eq:n=2lambdatilde2pt}
\langle \tilde{\lambda}_{h+1}^{-q}\left(z_{1}\right)\tilde{\lambda}_{h+1}^q\left(z_{2}\right)\rangle =\frac{\left(4h^{2}-q^{2}\right)\left(2ch+c-3\left(2h+q^{2}\right)\right)}{2ch-3q^{2}}\frac{1}{z_{21}^{2h+2}}.
\end{equation}

Using the decomposition (\ref{eq:Appendixn=2LambdaDecompose}), we can calculate  $\CV_{\lambda_L^{-q_L}\phi_L^{q_L}\phi_H^{-q_H}\phi_H^{q_H}}$,   $\CV_{\phi_L^{-q_L}\lambda_L^{q_L}\phi_H^{-q_H}\phi_H^{q_H}}$ and $\CV_{\lambda_L^{-q_L}\lambda_L^{q_L}\phi_H^{-q_H}\phi_H^{q_H}}$ from $\CV_{\phi_L^{-q_L}\phi_L^{q_L}\phi_H^{-q_H}\phi_H^{q_H}}$, and then equate these blocks to equations (\ref{eq:n=2philambdaphiphi}), (\ref{eq:n=2lambdaphiphiphi}) and (\ref{eq:n=2lambdalambdaphiphi}), to solve for $g_{2,h_L}(z)$, $g_{4,h_L}(z)$ and $g_{5,h_L}(z)$,  respectively. As we said in subsection \ref{Appendixn=2Leading}, at leading order of the large $c$ limit, these blocks only get contributions from Virasoro and $U(1)$ generators. Some details for calculating these blocks are as follow:
\begin{enumerate}
\item  In these calculations, we need to use the relationship between vacuum blocks with descendant fields and vacuum blocks with only primaries. These relationships are the same as those for the corresponding correlation functions, which are derived in subsection \ref{appendixn=2descendant}. 
\item  The heavy-light vacuum blocks with one light operator being $\tilde{\lambda}$ and the other light operator being $\phi$ vanish,
$\CV_{\tilde{\lambda}_L^{-q_L}\phi_L^{q_L}\phi_H^{-q_H}\phi_H^{q_H}}=\CV_{\phi_L^{-q_L}\tilde{\lambda}_L^{q_L}\phi_H^{-q_H}\phi_H^{q_H}}=0$. The reason is just because $\tilde{\lambda}_L$ and $\phi_L$ have different conformal dimensions ($h_{\tilde{\lambda}_L}=h_{\phi_L}+1$), and the two-point functions of them vanishes, $\langle \tilde{\lambda}_{L}^{-q}\phi_L^q\rangle=\langle \phi_L^{-q}\tilde{\lambda}_{L}^{q}\rangle=0$.
\item At leading order of the large $c$ limit, the only difference (up to normalization) between the vacuum blocks $\CV_{\tilde{\lambda}_L^{-q_L}\tilde{\lambda}_L^{q_L}\phi_H^{-q_H}\phi_H^{q_H}}$ 
and  $\CV_{\phi_L^{-q_L}\phi_L^{q_L}\phi_H^{-q_H}\phi_H^{q_H}}$ is that the conformal dimension of the light operators in the former is $h_L+1$ while that of the latter is $h_L$. So we can just change $h_L$  to $h_L+1$ in the expression of $\CV_{\phi_L^{-q_L}\phi_L^{q_L}\phi_H^{-q_H}\phi_H^{q_H}}$ to get\begin{equation}
\CV_{\tilde{\lambda}_L^{-q_L}\tilde{\lambda}_L^{q_L}\phi_H^{-q_H}\phi_H^{q_H}}=\frac{\left(2 h_L+1\right) \left(4 h_L^2-q_L^2\right)}{2 h_L}e^{(h_L+1)\tilde{f}(z)}(1-z)^{-3\eta_q q_L},
\end{equation}
where the prefactor here is just the prefactor in (\ref{eq:n=2lambdatilde2pt}) in the large $c$ limit. 
\end{enumerate} 


\bibliographystyle{utphys}
\bibliography{VirasoroBib}

\end{document}